\documentclass[12pt,preprint]{aastex}

\newcommand\teff{T_{\rm eff}}
\newcommand\fsed{f_{\rm sed}}

\shorttitle{Evolution of L \& T Dwarfs in CMDs}
\shortauthors{Saumon \& Marley}

\begin{document}

\title{The Evolution of L and T Dwarfs in Color-Magnitude Diagrams}

\author{D. Saumon}
\affil{Los Alamos National Laboratory, PO Box 1663, Mail Stop F663, Los Alamos, NM 87545}
\email{dsaumon@lanl.gov}

\author{Mark S. Marley}
\affil{NASA Ames Research Center; Mail Stop 245-3; Moffett Field CA 94035}
\email{Mark.S.Marley@NASA.gov}

\begin{abstract}
We present new evolution sequences for very low mass stars, brown dwarfs and giant planets and 
use them to explore a variety of influences on the evolution of these objects.  While the 
predicted adiabatic evolution of luminosity with time is very similar to results of previous work, 
the remaining disagreements reveal the magnitude of current uncertainty in brown dwarf evolution theory.
We discuss the sources of  those differences and argue for the importance of the surface 
boundary condition provided by atmosphere models including clouds. 

The L- to T-type ultracool dwarf transition can be
accommodated within the \citet{am01} cloud model by varying the cloud sedimentation parameter.
We develop a simple model for the evolution across the L/T transition. 
By combining the evolution calculation and our atmosphere models, we generate colors and magnitudes of
synthetic populations of ultracool dwarfs in the field and in galactic clusters. We focus on near infrared
color-magnitude diagrams (CMDs) and on the nature of the ``second parameter'' that is responsible for
the scatter of colors along the $\teff$ sequence.  Instead of a single second
parameter  we find that variations in metallicity and cloud parameters, unresolved binaries and
possibly a relatively young population all play a role in defining the spread of brown dwarfs along the 
cooling sequence.  We also find that the transition from cloudy L dwarfs to cloudless T dwarfs
slows down the evolution and causes a pile up of substellar objects in the transition
region, in contradiction with previous studies.
The same model is applied to the Pleiades brown dwarf sequence with less success, however.  Taken at
face value, the present Pleiades data suggest that the L/T transition occurs at lower $\teff$ for lower gravity
objects, such as those found in young galactic clusters.  The simulated populations of brown dwarfs also
reveal that the phase of deuterium burning produces a distinctive feature in CMDs that should be
detectable in $\sim 50$--100$\,$Myr old clusters. 
\end{abstract}


\section{Introduction}

There are now approximately 450 L dwarfs and 100 T dwarfs known (see \citet{kirk05} for a review of these spectral classes).  
They span effective temperatures from about 2400 to 700 K and exhibit a range of gravities, metallicities, 
and atmospheric condensate contents.  After more than a decade of intense study,
the modeling of the complex atmospheres and synthetic spectra of brown dwarfs
has reached a rather high degree of sophistication, including the chemistry of a very large number
of gas and condensate species \citep{allard01,lod02}, increasingly complete molecular opacity databases \citep{fml08,sb07}, extreme 
resonance line broadening \citep{bms00,aak05}, and particulate cloud models \citep{allard01,am01,tsu02,helling08}.  While a 
few conspicuous problems remain, the synthetic spectra and colors reproduce the observations fairly well and the 
determination of the basic astrophysical properties of brown dwarfs has begun in earnest.
The full astrophysical benefit of synthetic spectra and colors is obtained when atmosphere calculations
are coupled with evolution models that provide the surface atmospheric parameters $(\teff,g)$ as
a function of mass and age. The time evolution of the spectrum and colors, as well as absolute
fluxes can be computed and directly compared with observations to estimate astrophysical parameters
that are not easily amenable to direct observation.

To enable such comparisons using our own model atmosphere effort and to pursue a more complete analysis 
of spectroscopic and photometric data,
we have developed a code to compute evolution sequences of low mass stars, brown dwarfs and giant
planets.  These evolution sequences have already been applied extensively to the
analysis of brown dwarf observations \citep{roellig04,saumon06,saumon07,leggett07y,
leggett07p,cushing08} but have not been discussed in any detail.  
Here, we describe the input physics and assumptions of our particular approach as 
well as the unique aspects of our atmospheric boundary condition.  
The evolution of isolated brown dwarfs---a relatively
simple case of stellar evolution---has been studied extensively for the past 20 years
and is well understood \citep{dm85,nrj86,bhl89,bl93,bur97,cbah00a,cbah00b,bcah02}.  
Our evolution model is similar to the more recent work and the result are very similar
to previous work. A detailed comparison with other published calculations of the 
evolution of brown dwarfs reveals small differences that we quantify and, to the extent possible,
attribute to the different assumptions and approximations in each model.
We highlight the application of the evolution
sequences to the calculation of synthetic near infrared color-magnitude diagrams (CMDs) to 
explore the nature of the ``second parameter'' responsible for the spread of observed objects
around the main trends along the L and T spectral sequences. 
We develop a simple parametric model for the L/T transition to reproduce with good success the
CMD of field brown dwarfs.  We apply the same model to synthesize the population of the
Pleiades cluster (110$\,$Myr) and compare with
the latest deep survey data for this galactic cluster.  We discuss several potentially observable
features in near-infrared CMDs that would illuminate the evolution of brown dwarfs as well as
the nature of L/T transition.

\section{Evolution model: Assumptions and physical inputs}
The evolution model assumes adiabatic cooling of spherical, hydrostatic, non-magnetic,
non-rotating brown dwarfs.  The adiabatic assumption is valid for dense fully convective
structures and objects with masses ranging from $0.3\,M_\odot$ down to Saturn's mass can be modeled. 
In this section we consider key physical aspects of the evolution model.

\subsection{Electron Conduction}

The effects of electron conduction on the cooling of old brown dwarfs, pointed out by 
\citet{cbah00a}, are not included in our adiabatic calculation.  Conduction becomes significant 
only for the more massive brown dwarfs and at ages greater than about 2 Gyr \citep{cbah00a}.
Because the structure of a brown dwarf model is determined primarily by the 
pressure of the degenerate electrons but 
its heat content is given by the temperature of the ions, 
a cooling brown dwarf goes through very nearly the same sequence of $L,\teff ,R$, but
at later times when heat transport by conduction is included. 
We will see in \S 4 that neglecting conduction produces a relatively modest error when
considering other sources of uncertainty.

\subsection{Thermal Structure and evolution}

The adiabatic evolution is obtained by solving the mass equation
\begin{equation}
{dm \over dr}=4\pi \rho r^2,
\end{equation}
the equation of hydrostatic equilibrium
\begin{equation}
{dP \over dr}= -{\rho G m(r) \over r^2},
\end{equation}
and the equation of conservation of energy
\begin{equation}
\Bigg( L -\int_0^M \epsilon_{\rm nuc}\,dm \Bigg) dt=-\int_0^M TdS\,dm,
\end{equation}
where the rate of nuclear energy generation is $\epsilon_{\rm nuc}(\rho,T)$ and the equation of
state (EOS) along an adiabat is $P(\rho,S)$ and $T(\rho,S)$.  In these equations, $\rho$, $P$, $T$ and
$S$ are  the density, the pressure, the temperature, and the entropy of the gas, respectively,
$L=4\pi R^2\sigma\teff^4$ is the luminosity, and $m(r)$ is the mass interior to radius $r$.  The equation of conservation
of energy controls the evolution time scale as it gives the time step $dt$ between two successive adiabatic
structures that differ in entropy by an amount $dS$.  

We use the hydrogen and helium EOSs of Saumon, Chabrier \& Van Horn  (1995, hereafter SCVH), which
was developed specifically for this type of application.  This EOS is common to nearly all other
evolution and structure calculations of low mass stars, brown dwarfs, extrasolar giant planets,
Jupiter and Saturn (e.g. \cite{bhsl,bur97,cbah00a,cbah00b,bcah02,sg04}). 
Appendix A reports on typographical errors that were recently found in SCVH.  
The contribution of the metals to the EOS, which results in a slightly smaller radius for given 
mass and entropy, is neglected.

\subsection{Nuclear Energy Generation}

Nuclear energy generation in low mass stars and brown dwarfs is quite simple and is
reduced to one branch of the pp chain \citep{bl93}:
\begin{equation}
p + p \rightarrow d + e^+ + \nu_e
\end{equation}
and
\begin{equation}
p + d \rightarrow ^3{\rm He} + \gamma.
\end{equation}

We use the nuclear reaction cross-sections
from the NACRE data base \citep{nacre} and apply the screening corrections developed
by Chabrier which include both ionic and electronic screening (Saumon et al. 1996;
Chabrier \& Baraffe 1997; Chabrier, priv. comm.).  We assume that the compositional
changes due to nuclear reactions are homogenized by convection throughout the brown dwarf on a time scale much shorter
than the nuclear burning time scale (see Chabrier \& Baraffe 1997, however).  Lithium burning
is a useful age/mass diagnostic of brown dwarfs \citep{rmm92,basri00} but contributes negligibly to
nuclear energy production and is not included in our code.

\subsection{Initial State}

The initial state for the evolution assumes an extended, high entropy, spherical configuration
defined by $\teff>4000\,$K (higher for larger mass).  Such an idealized initial
condition bears little relation with the actual formation of brown dwarfs.
\citet{bcah02} have shown that the evolution of brown dwarfs is quite
sensitive to the choice of initial state for ages under $\sim 1\,$Myr.  For this reason,
we present only the evolution for later times.  The early evolution of our models
is further limited by the high-temperature limit of the atmosphere grid (see below).
For the initial interior composition, we adopt $Y=0.28$, $Z=0$,  and D/H$=2\times 10^{-5}$ by number
($2.88\times 10^{-5}$ by mass), and we set the abundance of $^3$He to zero.  

\subsection{Surface Boundary Condition}

The most significant difference between the BD evolution calculations published over the past
decade is in the treatment of
the surface boundary condition of the model, which connects the surface properties
($\teff,g$) to the interior model ($M$, $L$, $R$, age). 
For our adiabatic (i.e. constant entropy) interior models, applying the surface boundary condition
consists of matching the interior entropy to that of the convective bottom 
of the atmosphere model.  The importance of using
realistic, non-gray atmosphere models for the surface boundary condition of
very-low mass stars and brown dwarfs has been emphasized for some time \citep{saumon94,cb97}.
We use our grids of atmosphere models 
to obtain the surface boundary condition which is expressed as the entropy at the bottom of
the atmosphere as a function
of the atmospheric parameters, $S(\teff,g,{\rm [M/H]},f_{\rm sed})$, where $f_{\rm sed}$ is a cloud model
parameter (see below).  This makes the evolution fully self-consistent with the modeled spectra.

Our atmosphere models have been described previously \citep{mckay89,marley96,bur97,mm99,mar02} and used in several  detailed
comparisons with data \citep{marley96, mar02, roellig04, saumon06, saumon07, leggett07p, mainzer07, blake07, cushing08}.  
An updated, detailed description will be the subject of an upcoming publication. Briefly, the model computes radiative-convective equilibrium
atmospheric temperature-pressure structures.  Our thermal radiative transfer follows the source function technique of \citet{toon89}
allowing inclusion of arbitrary Mie scattering particles in the opacity of each layer.
Our opacity database, accounting for all important absorbers is described in \citet{fml08}. 
Our chemical equilibrium grid of molecular, atomic and ionic
abundances as a function of temperature, pressure, and metallicity is based on the work
of \citet{fl94,fl96,lod99,lod02,lf02} and \citet{lf06}.  We use the elemental abundances of \citet{lod03}.
Our baseline cloud model \citep{am01} parametrizes the efficiency of sedimentation of cloud particles
through an efficiency factor, $f_{\rm sed}$. Large values of $f_{\rm sed}$ correspond to rapid particle growth and
large mean particle sizes. In this case condensates quickly fall out of the atmosphere, leading to
physically and optically thin clouds.  When $f_{\rm sed}$ is small, particles grow more slowly
and the atmospheric condensate load is larger and clouds thicker.  
Our cloud model is fully coupled with the radiative transfer and the $(P,T)$ structure of the model
during the calculation of a model so that they are fully consistent when convergence is obtained.
The model has been very successful
when applied to the atmospheres of brown dwarfs and Jupiter \citep{am01, knapp04,golim04}.
In particular Marley et al. (2004) and \citet{cushing08}
show generally excellent fits between our model spectra and observations of cloudy L dwarfs.
The near infrared colors of brown dwarfs are quite sensitive to the choice of $f_{\rm sed}$,
a point we will return to in the discussion of color-magnitude diagrams (\S 4).

Our grid of atmosphere models
covers $500 \le \teff \le 2400\,$K and $3.5 \le \log g \le 5.5$, which does not provide a boundary
condition for
the late evolution of low-mass objects and the very early evolution of the more massive ones.
To obtain a boundary condition for $\teff < 500\,$K, we define $T_{10}$ as
the temperature at $P=10\,$bar that gives the same entropy $S$ as the atmosphere model
\citep{lunine89}. We smoothly interpolate $T_{10}(\teff,g)$ quadratically to $T_{10}=0$
at $\teff=0$ at constant gravity. The interpolated values of $T_{10}$ are then converted back to
an equivalent entropy $S(T_{10},P=10\,{\rm bar})$ for use as the surface boundary condition.
For $\teff>2400\,$K and
gravities outside the range of the atmosphere grid, linear extrapolation in $\teff$
and $\log g$ are used.  This gives well-behaved but inaccurate results for the 
early, warm phase of the evolution and for very-low mass objects ($M\lesssim 0.002\,M_\odot$).  

In the following we present two evolution sequences, one based on a cloudless
atmosphere grid and one using a cloudy grid with $f_{\rm sed}=2$, both with solar
metallicity.  The two surface boundary conditions (expressed as $T_{10}$)  are shown in Fig.
\ref{fig:BC} for selected gravities.  Clouds
play a minor role in high-$\teff$ atmospheres and both sequences converge to the same
boundary condition.  The cloudy surface boundary condition is not as smooth as
the cloudless case because of
occasional numerical difficulties in the cloudy atmosphere calculation.
Cloudy atmospheres have a globally larger opacity than clear models, which results
in higher entropy in the convection zone (i.e., a hotter interior structure, or higher $T_{10}$, Fig. \ref{fig:BC}).
The differences between the cloudy and cloudless surface boundary conditions are significant.  
Calculations of the evolution of cloudy models with other values of $f_{\rm sed}$ are possible
and desirable but require an increasing number of model atmospheres to define the surface boundary condition.
The boundary conditions from restricted grids of models for $f_{\rm sed}=1$ -- 4 shows less than half of the 
difference in $T_{10}$ that is seen between $f_{\rm sed}=2$ and cloudless models, and  even less in
some regimes. To a good approximation, the $f_{\rm sed}=2$ evolution sequence can be used for
all cloudy cases.

\section{Evolution sequences}

\subsection{Cloudless and cloudy surface boundary conditions}

The evolution of the luminosity is shown in Fig. \ref{fig:evo_nc} for cloudless
models and Fig. \ref{fig:evo_f2} for the cloudy models for objects ranging
from 0.005 to 0.08$\,M_\odot$. Both show the well-known features of the evolution of very low mass stars, brown dwarfs and
giant planets \citep{dm85,bhl89,bur97}, with an initial deuterium burning phase
($M\gtrsim 0.015\,M_\odot$), followed by contraction and eventual settling on the
main sequence for the higher mass models ($M \gtrsim 0.075\,M_\odot$). The effect of
atmospheric opacity on the cooling of brown dwarfs is also well known
\citep{bhsl,bl93,cb00,cbah00a} and a comparison of the cloudy and cloudless sequences
shows the expected trends.  The cooling rate is primarily controlled by the slope
of the $T_{10}(\teff)$ relation (Fig. \ref{fig:BC}) and for $\teff \gtrsim 1000\,$K,
the cloudy models are cooler and less luminous for a given age \citep{cbah00a}.  
We find that at $\teff=1400\,$K, a cloudless brown dwarf is $\sim 35$\% older than
one with a $\fsed=2$ cloudy atmosphere, a value that is nearly independent 
of mass from 0.02 to 0.06$\,M_\odot$. 

For these evolution sequences with solar metallicity atmospheres, the minimum hydrogen-burning mass, defined
by objects that reach stable H-burning at an age of 10$\,$Gyr is 0.075$\,M_\odot$,
where $\teff=1910\,$K for the cloudless sequence.  The 
main sequence of cloudy models ends at a lower mass of $0.070\,M_\odot$ and $\teff=1550\,$K.  The latter falls
within the range where clouds have a maximal effect on the atmosphere and is
therefore a more realistic value of the hydrogen minimum burning mass of solar metallicity stars
than the cloudless case. Defining the
minimum deuterium-burning mass as objects that burn 90\% of their initial deuterium 
in 10$\,$Gyr, we find 13.1$\,M_{\rm Jupiter}$ and 12.4$\,M_{\rm Jupiter}$ for
the cloudless and cloudy cases, respectively.  

For comparisons with observables, a representation of the evolution
in terms of $\teff$ and gravity is more transparent (Figure \ref{fig:Teff_g_nc}).
The range of $\teff$  shown covers all spectral types later than L0, and includes
the bottom of the main sequence.  The figure also shows isochrones,
and lines of constant radius and luminosity.  The knowledge of any two quantities
among $\teff$, $g$, $R$, $M$, $L$ and age allows a determination of all the others. 
The most notable feature is a bump in all isochrones for masses between 0.01 and
0.02$\,M_\odot$ which signals the deuterium burning phase.  This figure shows
a number of interesting features of the evolution. For example, an object with a
known $L$ (red curves) has an upper limit to its $\teff$.  The figure also shows that there
is a maximum gravity for {\it any} brown dwarf in a sequence of models.  This arises from
the mass-radius relation of old, degenerate brown dwarfs which is primarily
determined by the equation of state.  For the cloudless
sequence of Fig. \ref{fig:Teff_g_nc}, the upper bound on the gravity is $\log g=5.465$, which 
occurs for a 0.068$\,M_\odot$
brown dwarf at $\teff=1160\,$K.  In the cloudy sequence, the peak gravity occurs at the same
mass but $\log g=5.366$ and $\teff=1380\,$K.  There is a similar weak dependence of
the maximum gravity on the metallicity, which increases by 0.023 dex for a decrease of 0.3 dex in
[M/H] (for cloudless sequences).  More generally, there is an upper ($\teff$,$\log g$) 
envelope to the evolution of brown dwarfs that in practice is defined by the main sequence and
the oldest objects ($\sim 10\,$Gyr for the disk population).
This envelope puts a firm upper limit
on determinations of the gravity of brown dwarfs (for a given $\teff$) such as those obtained by fitting spectra.
Higher values of the gravity are unphysical.

The ($\teff$,$\log g$) evolution of the cloudy sequence is shown in Fig. \ref{fig:Teff_g_f2} where
the cloudless cooling tracks are also shown for comparison.  The effects of clouds
on the surface boundary condition (Fig. \ref{fig:BC}) essentially vanish at high and low
$\teff$, and consequently, the cooling tracks of the cloudy models overlap  the
cloudless tracks in both limits.  At intermediate $\teff$, where the cloud effects are
maximal, the gravity of an object of a given mass is up to 0.08$\,$dex lower in the cloudy models than for the cloudless case.
For given values of ($\teff$,$\log g$), differences in $\log L$
can range for 0 to 0.1 dex, from 0 to 6\% in radius, and 0 to 60\% in age. These differences 
are modest but will become significant as brown dwarf studies become increasingly detailed and
precise. 

\subsection{Comparison with other calculations}

We validate our calculation by comparing with published results that are
widely cited in the brown dwarf literature.  This also offers a measure of the relative
uncertainties that remain in modeling the evolution of brown dwarfs and, to some extent, to
identify the model assumptions responsible for those differences.
Figures \ref{fig:evo_nc} and \ref{fig:evo_f2} compare our cloudless sequence with the COND03 models of
\citet{bcbah03}  and our cloudy $f_{\rm sed}=2$ sequence with the DUSTY00 models
of \citet{cbah00a} and \citet{bcah02}, respectively.  
We note that the COND03 and DUSTY00 models and ours use the same H/He EOS, the same interior composition
and that there is no significant difference in our respective nuclear reactions rates
(including screening factors).  Except for the modest role of electron conduction, 
the differences can only arise from the initial
boundary condition (for ages under a few Myr) and the surface boundary condition
provided by the atmosphere at later times.

Generally, the COND03 and DUSTY00 luminosities at a given age are slightly higher than
ours.  In both cases, however, the agreement is excellent for all masses
at all ages above a few Myr, with differences below 0.05 dex for the cloudless case and
below 0.09 dex (and usually much less) for the cloudy sequence.  At young ages, the 
differences can be significant and
reflect our different choice for the initial conditions for the models \citep{bcah02}.
For massive brown dwarfs at late times, electron conduction becomes the dominant
energy transport mechanism in the core (\S2.1).  At 10$\,$Gyr of
age, the luminosity of the COND03 models $\lesssim 0.06\,M_\odot$ is systematically higher than 
ours, an effect that decreases for lower masses.  This is the expected behavior for models
that include electron conduction. For a 10$\,$Gyr old brown dwarfs of 0.06$\,M_\odot$, we find
a luminosity that is 0.104 dex lower than that of \citet{bcbah03}, which is in perfect
agreement with the value given in \citet{cbah00a} for the change in luminosity brought
by introducing conduction in the evolution.  Thus, the differences at late times are primarily due
to our neglect of conduction and those differences are rather small.

A comparison of our cloudless models with the evolution sequence of \citet{bur97} (hereafter, B97)
is shown in Fig. \ref{fig:evo_B97}.
The agreement is not as good as we found with the COND03 models (Fig. \ref{fig:evo_nc}).
One of the reasons is that the B97 models use a rather low value for the
helium abundance of $Y=0.25$, while the present models and the COND03 models use $Y=0.28$.
The proto-solar value is $Y=0.274\pm0.12$ \citep{lod03}.  To better compare with B97, we have computed a short sequence
of models with $Y=0.25$.  This systematically shifts the luminosity curves downward
and reduces the differences but not enough to reach the level of agreement  seen in Fig. \ref{fig:evo_nc}.  
The B97 models also use the same EOS and nuclear reaction rates as in our calculation.  The
screening correction \citep{graboske73} is more crude but this only affects models at the edge of
the main sequence.  The primary reason for the differences, and for the structure seen
in the B97 $L(t)$ curves at $\log L/L_\odot \sim -4$ that is not apparent in the COND03 or in our cloudless sequence is the 
surface boundary condition. The model atmospheres used in B97 are the non-gray cloudless models
of \citet{marley96}, complemented with gray cloudless models at the higher $\teff$ \citep{saumon96}. 
Atmosphere models of brown 
dwarfs have become increasingly more realistic over the past decade.
Particularly significant changes in cloudless models since the work of \citet{marley96}
are 1) the recognition of the role of the K I and Na I resonance doublets at
optical wavelengths \citep{bms00}, 2) improved H$_2$ CIA opacity \citep{bjf01,borysow02}, 3)
a new line list for TiO \citep{ahs00,fml08},
4) improved modeling of the condensation chemistry \citep{lod02}, 5) new sources of
molecular opacity such as CrH, FeH, VO and PH$_3$ (\citet{fml08} and references therein), 
and 6) expanded molecular opacity lines list -- notably for CH$_4$
and NH$_3$ \citep{fml08}.  While the evolution is not sensitive to the details of the atmospheric opacity,
the global opacity of the atmosphere
largely accounts for the differences between the various evolution calculations for
brown dwarfs.  The above factors and cloud formation as well, all play a significant role in increasing
the atmospheric opacity, each in a different range of $\teff$ and gravity. The analytic evolution of \citet{bl93}
show that the luminosity increases with the Rosseland mean atmospheric opacity as $L\sim \kappa_{\rm \scriptscriptstyle R}^{0.35}$, a
relation confirmed by our non-solar cloudless evolution sequences.  This, along with the choice of $Y=0.25$
explains most of the differences between our calculation and that of B97.

\section{Synthetic color-magnitude diagrams}

The availability of parallaxes for brown dwarfs \citep{dahn02,tbk03,vrba04} provides absolute magnitudes and spectral fluxes 
that can be compared with a combination of evolution models and synthetic spectra and colors.
The near infrared CMD of field brown dwarfs (Fig. \ref{fig:CMD_data}) shows a number of remarkable features.
The L dwarf sequence naturally extends the M dwarf main sequence, where evidence for condensates in the atmosphere first
appears \citep{jt97} suggesting that clouds play a continuous role from late M dwarfs through the L spectral sequence.
The photometric sequence is increasingly broader with later spectral types, up to the latest L dwarfs, which
suggests the presence of a ``second parameter'' besides $\teff$ that affects the colors of L dwarfs.
\citet{tsu03}, \citet{knapp04} and \citet{bsh06} have suggested that variations in gravity (i.e. a spread in mass and age)
account for the broadening.  Other factors such as metallicity, unresolved binaries, and a range of cloud properties
may also contribute to the scatter.
Between the latest L dwarfs and the mid-T dwarfs, there is a dramatic shift of $J-K$ to the blue, long recognized as due to 
the lack of cloud opacity in late T dwarfs and the appearance of CH$_4$ in the $K$ band. The late T dwarf sequence is also fairly broad, providing
further evidence
of the importance of additional physical parameters in brown dwarf atmospheres.  An important revelation enabled by CMDs is 
that mid-T dwarfs are brighter in $J$ (but not in $H$ or $K$) than the latest L dwarfs 
\citep{dahn02,tbk03,vrba04}, even though the bolometric luminosity decreases steadily with later spectral types \citep{golim04}.  
This was confirmed by the discovery of binary brown dwarfs with (coeval) components spanning the L/T transition and showing the
$J$ band flux reversal \citep{burgasser06,liu06}.  Even more intriguing, the binary fraction
of early T dwarfs is twice as large as for other dwarfs and their spectra and colors are largely the
result of the combined light of components of earlier and later spectral types \citep{burgasser06,liu06}.  Together, these facts bear on 
the nature of the transition from L to T spectral types
which is believed to be caused by a rather dramatic change in the cloud properties (such as the fraction of cloud coverage,
particle size, and the cloud's vertical extent) over a small
range of $\teff$ of $\sim 200\,$K \citep{vrba04,golim04}. There is as yet no physical modeling of such a process.

Detailed comparisons of our self-consistent evolution and spectra with spectroscopic data of a few
late T dwarfs have been rather successful \citep{saumon06, saumon07}, demonstrating that at least for
the coolest brown dwarfs known the models, while admittedly imperfect, are fairly realistic.  Fits of the entire
spectral energy distribution of brown dwarfs with earlier spectra types are more challenging due to the cloud opacity but the
models have reached a level of realism where meaningful results can be obtained \citep{cushing08}. A comparison
of our model colors in the mid- and near-infrared with a sample of $\sim 50$ brown dwarfs 
shows generally good agreement \citep{leggett07p}.  The near infrared CMDs of our model colors are shown in
Fig. \ref{fig:CMD_atmos_MK} which displays the trends with $\teff$, gravity and the
cloud sedimentation parameter.  The $\fsed=1$ sequence generally follows the
observed L dwarf sequence and steadily becomes redder with decreasing $\teff$.\footnote{We note however that $\fsed=1$ models
produce relatively flat-topped $JHK$ band peaks, a feature not seen in spectroscopic data \citep{stephens08}.}
In models with decreasing cloud thickness
(increasing $\fsed$), the $J-K$ color is the same as for $\fsed=1$ at the high-$\teff$ end where the cloud opacity
is always modest but shows a turnover to
the blue that occurs at gradually higher $\teff$ as the thinner cloud deck sinks below the average photosphere.  
At low $\teff$, the colors of $\fsed=4$ models join those
of cloudless models ($\fsed \rightarrow \infty$) below $\sim 800\,$K. The cloudless models do not show the redward trend of the L sequence
but grow steadily bluer in near-infrared colors, due to the onset of H$_2$ collision-induced absorption and of CH$_4$ bands. 

In a very qualitative sense, the cloudy models show the turnover in $J-K$ seen in the L/T transition objects
but the turnover is much too gradual.  Furthermore, the cloudy models do not show the brightening in the $J$ band.
A comparison with Fig. \ref{fig:CMD_data}
indicates that {\it within the framework of the} \citet{am01} {\it cloud model}, the L to T transition 
can be modeled by increasing $\fsed$ during the cooling of a brown dwarf. This idea is supported by the detailed
analysis of the spectral energy distribution of several brown dwarfs \citep{cushing08}.
While it is plausible that the transition could be caused by an increase in sedimentation efficiency with
$\teff$ at the transition, the \citet{am01} model does not provide a physical justification for this process.
The transition could be the consequence of other cloud processes that are not captured in the Ackerman \& Marley
model. The rather simplistic cloud models developed so far are not adequate 
to understand the cause of the transition  and the complex dust formation model of \citet{helling08} has
not been applied to modeling the L/T transition.
Despite these caveats, the one-parameter Ackerman \& Marley cloud model reproduces observed spectra of
brown dwarfs of all spectral types \citep{cushing08} rather well and provides a useful basis for further
analysis of data.  Cloud models with more detailed physics, while weakly constrained
by present observations, are very much needed however.

Color-magnitude diagrams have historically been one of the most powerful tools to understand stellar evolution.
The difficulty in discovering and observing the intrinsically faint brown dwarfs has limited the
the application CMDs in this brand of stellar astrophysics.  Yet, a comparison
of modeled magnitudes and colors with a large volume-limited sample of brown dwarfs with measured parallaxes
could enable the determination of the cooling rate as a function of $\teff$, of the $\teff$ range of the
transition from cloudy to cloudless atmospheres, the mapping of the variation of cloud parameters with $\teff$,
and reveal the distributions of cloud parameters and gravities at a given $\teff$.  Such a sample
is not yet available. Furthermore,  the models are not quite reliable enough to carry out such a program.
On the other hand, the colors of synthetic populations of field brown dwarfs is a tool to explore the nature
of the ``second parameter'' along the L dwarf sequence  that complements the clues gathered from brown dwarf binaries.
In this section, we explore the CMDs of simulated brown dwarf populations based on our evolution sequences and
model atmospheres, compute a hybrid evolution sequence that models the L/T transition and make qualitative comparisons
with data in the $M_K$ vs $J-K$ CMD to identify trends under various assumptions. We first consider the population of
field brown dwarfs, followed by a discussion of brown dwarfs in the Pleiades.

\subsection{Disk brown dwarfs}

We simulate a volume-limited population of field brown dwarfs by generating a random set of points in the mass-age parameter space
with the mass chosen between 0.006 and 0.1$\,M_\odot$, assuming a power law initial mass function (IMF) 
\begin{equation}
{dN \over dM} = M^{-\alpha}
\end{equation} 
where $dN$ is the space density of brown dwarfs in mass interval $dM$.  We choose $\alpha=1$,
which is representative of the mass function of field brown dwarfs and consistent with the 
determination of \citet{allen05}.  
We adopt a constant star formation rate (SFR) over the age of the Milky Way (0 -- 10$\,$Gyr). 
The normalization of the number density $N$ is irrelevant for our purposes.  We arbitrarily normalize all
simulations to a fixed number of objects in the mass range 0.075--0.08$\,M_\odot$.  
Because all brown dwarfs with known parallax
are nearby, we neglect extinction and reddening.  For simplicity, our synthetic population consists of
single dwarfs only.  The effects of binaries and of different assumptions for the SFR and the IMF are discussed separately in \S4.2.1.

We convert our sample of ($M$, age) points to ($\teff, \log g$) values with our evolution
sequences.  This is shown in Fig. \ref{fig:disk_distrib} where we have used the cloudy evolution sequence
with $f_{\rm sed}=2$.  Objects with $\teff$ above the upper limit of our surface boundary condition (2400\,K) have
been culled from the sample.  The distribution can be compared with the cooling tracks and isochrones for this sequence (Fig. \ref{fig:Teff_g_f2}).
The resulting ($\teff$, $\log g$) distribution is rather remarkable. As can be expected for a population with an
average age of 5$\,$Gyr, the vast majority of objects are concentrated near the outer envelope of
the evolution defined by the 10$\,$Gyr isochrone.  The upper boundary, between $\log g=5.3$ and 5.4,
contains a narrow distribution formed by the low-mass end of the main sequence and the most massive
brown dwarfs ($M \gtrsim 0.06\,M_\odot$).  The vast majority of objects are found at $\teff \lesssim 1000\,$K;
a consequence of the rising IMF at low masses and of their rapid cooling.  
Changing the IMF power-law index $\alpha$ alters the relative abundance
of points along the low-$\teff$ edge of the distribution but the general appearance of this figure is
unaffected:  A thin distribution at high gravities and a broader, more populous distribution at low-$\teff$.
Most of these objects are very dim, however, as $\log L/L_\odot  \lesssim -6$ for
$\teff \lesssim 600\,$K.  Figure \ref{fig:disk_distrib} illustrates that warm, low gravity brown 
dwarfs would be quite rare in the field under the assumption of a constant SFR rate over the age of
the Galaxy.  In contrast, the kinematics of L and T dwarfs indicates that the population may be as young
as 0.5--4$\,$Gyr \citep{zo07}, a possibility we consider in \S4.2.1.

The pioneering work of \citet{chabrier02} and the extensive studies of \citet{burgasser04}, \citet{allen03} and 
\citet{allen05} show how variations in the IMF and SFR 
affect one-dimensional distribution functions (such as the luminosity function) of field brown dwarfs.  
They offer a clear and detailed discussion of the features of the very-low mass stellar and substellar
luminosity function in terms of the characteristics of their evolution. These studies
focus primarily on a determination of the substellar mass function and the mass budget of the Galaxy \citep{chabrier01,chabrier02}
or on the luminosity function which collapses all objects from a two-dimensional distribution
-- for example, age and mass -- into a one-dimensional distribution, binned in terms of $\teff$, $L$, or
spectral type.  Our primary concern is to explore the two-dimensional distribution and its potential to reveal additional features
of the evolution of brown dwarfs by constructing color-magnitude diagrams.  Figures (\ref{fig:panels_fsed}) show
our synthetic field distribution of brown dwarfs in the ($M_K$, $J-K$) CMD for different values of the
cloud condensation parameters $f_{\rm sed}$.  All four figures are based on the same cloudy evolution
sequence with $f_{\rm sed}=2$ and the ($\teff,g$) distribution of Fig. \ref{fig:disk_distrib}, but each uses 
colors computed from atmosphere models with different values of $\fsed$. 
An evolution sequence with a single value of $\fsed$ should  be adequate for
most purposes (see \S2.5).  Finally, Fig. \ref{fig:panels_fsed}e shows the cloudless case, computed with
the cloudless evolution sequence and cloudless atmosphere models.  The figures also shows the photometry of
M, L, and T dwarfs with known parallaxes, excluding known binaries with blended 
photometry\footnote{For brevity, we do not show the other near infrared CMDs that can be generated with the $JHK$ band passes.  The general
appearance of the other diagrams is very similar to that of $M_K$ vs $J-K$ as far as the comparison between the modeled
population and the data is concerned.}.  We note that since the set of dwarfs with measured parallaxes 
is not in any sense a volume-limited sample and is subject to unquantifiable biases, the comparisons must be regarded with caution.
For the same reasons a detailed, quantitative comparison of the modeled and observed two-dimensional distributions would be premature
and our study is mostly qualitative in nature. 

The most striking feature in these diagrams is that the color-magnitude distribution
of the sample is rather different from that of the simulation, the former being more or less evenly weighted between L  and T dwarfs 
while the simulation predicts that T dwarfs should be dominant if the IMF keeps rising towards low masses \citep{allen05,burgasser04}.  
This shows that the sample of brown dwarfs with measured parallaxes is more similar to a magnitude limited
sample than the volume limited sample we simulated \citep{chabrier02}.  We now turn to a comparison of our synthetic CMDs with
the observed features, starting with the high-$\teff$ end of the sequence and proceeding to the cooler brown dwarfs. 
Simulations under different assumptions are discussed in \S4.2.1.

At the upper limit of our calculation ($\teff=2400\,$K), the synthetic sequence joins the late M sequence nicely.
The cloudless and cloudy sequences meet at $\teff\sim 2400\,$K as clouds play a minimal role at higher $\teff$ (Fig. \ref{fig:CMD_atmos_MK}).
In the synthetic population, the $J-K$ color of the objects with $10.5 \lesssim M_K \lesssim 11.5$ 
becomes too blue by $\sim 0.3$, a problem that has plagued low temperature atmosphere models for several years. We find that
the effect is less pronounced in $J-H$, where the difference is $\sim 0.1$.
More extensive calculations of the TiO and H$_2$O line lists, while substantially improving the agreement with observed
spectra at optical wavelengths, have both increased the discrepancy with the observed $J-K$ color \citep{ahs00,bcah01}. 
Since TiO and H$_2$O are the dominant absorbers in the optical and near infrared, 
remaining limitations in the TiO \citep{ahs00}, and more likely, the H$_2$O \citep{ps97} line lists at high temperatures 
could reasonably be at the origin of this problem. On the other hand, dust opacity becomes significant in this
regime and reddens the near infrared colors (Fig. \ref{fig:CMD_atmos_MK}).
At the $\teff$ of late M dwarfs, the surface mass density of condensible material
is modest, but the opacity of a tenuous cloud can be increased by decreasing the size of the grains.  It is also possible 
that the onset of cloud formation in late M dwarfs occurs at higher $\teff$ than our models predict, as originally 
suggested by \citet{tsuji96}. The CMD suggests that the models may be underluminous rather than too blue
(or a combination of both).  The mass-radius relation for low mass stars is well-reproduced by evolution models down to
$\sim 0.096\,M_\odot$ \citep{cb00}, but there is growing evidence that models of late L dwarfs are systematically
underluminous for a given mass and age, as revealed by the first few dynamical mass measurements below
the stellar mass limit \citep{ireland08,dupuy08}.  For late T dwarfs, absolute model fluxes agree very well with observed spectral energy
distributions \citep{saumon06, saumon07} so the theoretical radii of BDs older than $\gtrsim 1\,$Gyr  and $\teff \lesssim 900\,$K appear
reliable.  Finally, there are indications that the nearby field brown dwarfs may be a relatively young population \citep{allen05,mh06,ldi08,cruz08}, 
which would decrease the typical gravity compared to our simulation, increase the effect of clouds and move the population towards 
redder $J-K$ (Fig. \ref{fig:CMD_atmos_MK}).  
The discrepancy in the early L sequence could be the combination of a number of these effects. The input physics of the
evolution model is fairly well understood and the comparisons between three independent calculations show that the
results are fairly robust.  On the other hand, the colors of model (e.g. $J-K$) are sensitive to the details of the input
opacities of atmosphere models, which are known to be deficient in several ways. It appears more likely that the difference with
the observed sequence arises from deficiencies in the atmosphere models rather than in the evolution.

Small scale features in the synthetic CMDs are caused by the few cloudy atmosphere
models that have converged to a poor solution, such as can be seen in Fig. \ref{fig:CMD_atmos_MK}; 
they have no physical significance.  By construction, these CMDs fold in a distribution of gravities
and $\teff$ based on reasonable assumptions that can be compared with the observed distribution.  
The concentration of the more massive objects 
along a narrow line in Fig. \ref{fig:disk_distrib} remains clearly visible in the synthetic CMD. 
This feature is not apparent in the data and it may be blurred by reasonable variations in metallicity (Fig. \ref{fig:metal_f2})
The contribution of unresolved binaries to the blurring of the synthetic sequence is discussed in \S4.2.1. 
The breadth of the late M and early L
distribution is comparable to the scatter caused by variations in gravity, supporting the idea that the latter is indeed
the ``second parameter'' of the sequence \citep{tsu03,bsh06}.  

On the other hand, the widening of the observed distribution in $J-K$ of the mid to late 
L dwarfs cannot be explained solely by the range of gravity of the simulated population and appears to
require variations in cloud properties. A combination of $f_{\rm sed}=1$ and 2 populations 
appears sufficient to reproduce the observed range and the maximum $J-K$ of late L dwarfs.
This is confirmed by detailed spectral analysis of unusually red and blue $J-K$ colors with our synthetic spectra,
showing that L dwarfs span a range of $f_{\rm sed}=1-3$ \citep{burgasser08,cushing08,stephens08},
with the redder dwarfs having lower values of $f_{\rm sed}$.
Variations in metallicity increasingly affect
the $J-K$ color of late L dwarfs and could also contribute to the widening of the observed L sequence (Fig. \ref{fig:metal_f2}),
but probably not as much as variations in cloud properties.  Observations of 
binary L dwarfs such as the L4+L4 pair HD 130948BC \citep{dupuy08}, will present a set of spectral standards of known mass, age, 
and metallicity that will allow such effects to be disentangled. 

Our simulations indicate that a third parameter, related to cloud opacity,
is also necessary to model the distribution of the mid-L to mid-T dwarfs colors.  The transition from L to
T spectral types is thought to require a rapid change in the cloud properties over a 
small range of $\teff$.  The most important free parameter
in the \citet{am01} cloud model is the sedimentation efficiency $\fsed$.
All cloud properties (median particle size, vertical structure, etc.) follow self-consistently given
$\teff$, the gravity and the metallicity.  Within this model, the transition from cloudy L dwarfs to the
clear atmospheres of late T dwarfs can only be modeled by varying $\fsed$.  We find that the
late L dwarf sequence requires $\fsed=1$ to 2, and that
the transition dwarfs with $0.4 \lesssim J-K \lesssim 1.4$ are fairly well matched with
a $f_{\rm sed}=3$ cloud (Fig. \ref{fig:panels_fsed}c). Some parametrization of $\fsed$ as a function of $\teff$
(and perhaps gravity) could potentially match the CMD of L and T dwarfs.  Admittedly, a color-magnitude diagram that involves
only two band passes is a rather limited tool to study the transition.  Detailed fits of the entire spectral energy distributions of $\sim 20$ 
L1 to T6 dwarfs with these models fully support the present hypothesis as a trend of increasing $\fsed$ for T0 and 
later spectral types is found \citep{cushing08,stephens08}.

Finally, the $J-K$ colors of late T dwarfs are best reproduced by
our $f_{\rm sed}=4$ model as our cloudless models are somewhat too blue. 
Adding a thin cloud layer  makes the models slightly redder and brings them in better agreement with the data. 
Late T dwarfs are expected to be cloudless
however, as is indicated by detailed fits of their SED \citep{saumon06,saumon07, cushing08}. The overestimated blue color of our cloudless models
is most likely due to some inadequacy in the gas opacities, such as in the CH$_4$ line list or H$_2$ collision-induced absorption. 
In our models, the cloudless sequence is very narrow because the changes in $M_K$ and $J-K$ due to variations in gravity can be largely 
compensated by changes in $\teff$ (Fig. \ref{fig:CMD_atmos_MK}).  Therefore, variations in gravity cannot 
explain the scatter in the colors of late T dwarfs, in contrast with \citet{bsh06} who find that gravities
varying between $\log g=4.5$ and 5.5 are adequate for the task.  
Our result is essentially independent of the gravity dependence of
the modeled colors.  Late T dwarfs have $700 \lesssim \teff \lesssim 1200\,$K. The ($\teff$,$\log g$) distribution of
our $\alpha=1$ field population (Fig. \ref{fig:disk_distrib}) shows that the majority (87\%) of brown dwarfs in this
$\teff$ range have $\log g \ge 4.8$, with a distribution that peaks sharply at $\log g=5.3$.  Objects with $\log g \sim 4.5$ nearly 
all have cooled well below the $\teff$ of any object detected so far. Unless the sample of BDs with parallaxes contains
a large fraction of objects that are much younger than 5$\,$Gyr (the average age in this simulation), 
there are too few low gravity objects in the late T dwarf
$\teff$ range to account for the observed spread of the sequence, even when considering the different $JHK$ colors
of the cloudless sequence of \citet{bsh06}. 
It is conceivable that some late T dwarfs may not have completely cloudless atmospheres, which would increase the
variation in near-infrared colors of late T dwarfs. An example of such an object is 
the T5.5 dwarf SDSS J111009.99+011613.0 whose SED favors a model with $f_{\rm sed}=4$ \citep{stephens08}.
The dispersion caused by metallicity variations is discussed in \S4.2.1.

\subsection{A hybrid sequence of models}

To further illustrate the potential of population synthesis for the study of brown dwarfs, we have developed a
simple model of the L/T transition.  Based on the above discussion of the near-infrared CMD, we
consider an evolution sequence where the cloud sedimentation parameter $\fsed$ increases as
a function of decreasing $\teff$.  We 
linearly interpolate the surface boundary condition in $\teff$ (Fig. \ref{fig:BC}) between
cloudless models at 1200$\,$K and cloudy models ($\fsed=2$) at 1400$\,$K for each gravity.  For simplicity, 
we assume that the characteristics of the transition depend only on $\teff$.  The role of gravity in the physics of the
transition, if any, will have to be determined empirically.  We use the magnitudes of $f_{\rm sed}=4$
atmosphere models for $\teff \le 1200\,$K and of $f_{\rm sed}=1$ models above 1400$\,$K.
The slight inconsistencies between the $f_{\rm sed}$ values used in the evolution sequences and for 
computing the magnitudes are modest and can be ignored here (see \S 2.5).
The colors are linearly interpolated in $\teff$ at constant gravity between those two regimes.  Our 
linear interpolation between cloudy and cloudless models
implies a specific but arbitrary choice of the cloud evolution with $\teff$
across the L/T transition.  This approach to modeling the L/T transition is similar to that of
\citet{chabrier02} who interpolated magnitudes in $\teff$ between the DUSTY00 \citep{cbah00a}
and the COND \citep{cb00} evolution sequences and to \cite{burgasser02} who used a slightly more
elaborate interpolation between cloudy and cloudless magnitudes to approximate the effect of cloud
clearing at constant $\teff$.  Here we have recomputed the
cooling with the hybrid boundary condition which results in features in the evolution that
could not occur within the \citet{chabrier02} and \citet{burgasser02} schemes. 

The ($\teff, \log g$) distribution for this hybrid evolution sequence, under the same
IMF and SFR assumptions as in Fig. \ref{fig:disk_distrib} is shown in Fig. \ref{fig:disk_hybrid}. 
Two new features are apparent.  The most
apparent is a bump in the upper envelope of the distribution, where the maximum gravity rises
at the transition from the lower value of cloudy models to the higher value of cloudless models,
as discussed in \S 3.1.  This effect is modest ($\sim 0.1$ in $\log g$) but
is a direct consequence of the change in surface boundary condition
as the atmosphere becomes free of clouds.  In reality, this bump may be narrower or more spread out than 
is modeled here.  The other feature is an increased density of brown dwarfs in the 1400--1200$\,$K
range for $\log g\gtrsim 4.8$. The relatively high entropy interior (equivalently, higher $T_{10}$, see
Fig. \ref{fig:BC}) of a 1400$\,$K cloudy brown dwarf must release much more of its heat content to
become a 1200$\,$K cloudless brown dwarf than would normally be required for cooling to the same
effective temperature without 
the transition in cloud properties.  This increases the cooling time scale, i.e. the brown dwarf
cooling slows between 1400$\,$K and 1200$\,$K and causes a moderate pile up in the
$\teff$ range of the transition.  The effect can be clearly seen in the $\teff$ distribution of the
hybrid disk sequence where an excess of brown dwarfs of a factor of 2.1 occurs in the
transition region compared to either the cloudy or the cloudless evolution sequences (Fig. \ref{fig:Teff_distrib}).
For the same reason, an excess also occurs for the cloudy sequence relative to the cloudless case for 
$650 \lesssim \teff \lesssim 1000\,$K as the cloud sinks in the atmosphere to levels where it no longer
affects the surface boundary condition (Fig. \ref{fig:BC}). This is a more gradual effect than
our modeled L/T transition and the excess brown dwarf density is spread over a wider range of temperatures.
Qualitatively, this effect is an inevitable consequence of the disappearance of clouds at the
L/T transition and is not dependent on any particular physical mechanism or model of the
``cloud collapse'' responsible for the transition.  

This excess of brown dwarfs in the transition is in apparent contradiction with \citet{burgasser07} who found
a minimum in the space density of transition dwarfs as a function of spectral type.  There are two reasons
for this.  Our simulation of cloudy and cloudless evolution indicates a monotonous increase in the $\teff$ distribution towards lower
temperatures. As pointed out by \citet{burgasser07}, once converted to a distribution in terms of spectral type 
(or absolute magnitude) the distribution would show a minimum because the 
relation between $\teff$ and spectral type is highly non-linear and shows
a plateau in the L/T transition region \citep{golim04}.  The other effect is the slowing of
the evolution of brown dwarfs in our hybrid evolution sequence as the brown dwarf converts from a
cloudy to a cloudless atmosphere. To the best of our knowledge, this effect is not included in any of
the earlier population syntheses which are based on a single evolution sequence (cloudy or cloudless).

Also shown on Fig. \ref{fig:cluster_hybrid} are $(\teff, \log g)$ determinations for various field and 
ultracool dwarfs in binaries.  Given the small number of constrained objects and the current lack of detections at 
very low $\teff$, the agreement between the modeled and observed populations is reasonable.  As spectral fitting becomes more precise,
particularly for cloudy L dwarfs and for binary brown dwarfs, and with the creation of a larger, volume-limited
sample of BDs with parallaxes, such comparisons will test our understanding of ultracool dwarf formation and evolution.

The corresponding CMD is shown in Fig. \ref{fig:CMD_disk_hybrid}. The general agreement with the
data is quite good as we have chosen the parameters  of the hybrid model sequence for that purpose.
Figure \ref{fig:CMD_disk_hybrid} is to be compared to Fig. \ref{fig:panels_fsed}a for $\teff \ge 1400\,$K
and Fig. \ref{fig:panels_fsed}d for $\teff \le 1200\,$K.
The region of interest is the transition region where the synthetic population shows a well-defined
lower envelope whose shape and location depend on the details of the transition process. The
existence of this lower boundary simply reflects the high gravity envelope of Fig. \ref{fig:disk_hybrid}.
There is increased scatter in the transition region that arises
from the distribution in gravity and cloud sedimentation parameter. The larger density of brown dwarfs 
at $\teff \sim 1300\,$K seen in Fig. \ref{fig:disk_hybrid} is also visible in the CMD, centered
at $J-K \sim 1$.  The clump of five early T dwarfs is superposed on this density enhancement in the 
synthetic CMD but this is not a statistically significant result.  This subtle feature may be discernible in a much
larger data set, however.  Adaptive optics imaging of early T dwarfs has resolved many into earlier and later components, 
effectively depopulating the transition region \citep{liu06,burgasser06}. Nonetheless, the cooling of brown dwarfs is
a continuous process and the full range of $\teff$ must be populated regardless of the distribution as a function of spectral
type.  Our synthetic CMD shows that there should be
no significant dearth of objects in the transition region ($J-K \sim 0.2$ -- 1.5).  The current paucity of known transition
objects must be a selection effect.
The above remarks also apply to the $M_J$ vs $J-K$ CMD  where we obtain an equally
good agreement with the data, including the observed brightening in the $J$ band. The $M_H$ vs $J-H$ CMD shows the same
general features but with a larger dispersion below the transition. 
While the transition can be modeled with success by varying $\fsed$ in our atmosphere models, we do not propose
that this is the physical mechanism of the transition.  It is merely a parametric description
of the transition using a specific (and simple) cloud model. 

\subsubsection{Synthetic CMD with different assumptions}

In this section, we consider different assumptions for our population synthesis with the hybrid evolution
sequence to estimate the importance of variations in the SFR, the IMF, and the role of unresolved binaries.
These experiments are summarized in Fig. \ref{fig:panels_variations} where our fiducial
simulation of Fig. \ref{fig:disk_hybrid} is reproduced in panel {\it a} for convenience. Since there is
evidence that the local BD population has a younger average age than solar-type stars \citep{allen05,mh06,zo07,ldi08,cruz08}, 
we consider a SFR where all BDs were formed at a constant rate in the past 5$\,$Gyr (Fig. \ref{fig:panels_variations}b). This results in a larger fraction
of low-gravity objects, widening the spread of the sequence all the way to $J-K \sim 0$. The sharp blue edge of the sequence remains
well-defined above $M_K=12$ but the wider sequence reproduces the data fairly well.  If indeed, the local population of BDs turns out
to be younger than our fiducial SFR would imply, this would reduce the contribution of cloud properties as the ``second parameter'' 
favoring the age (gravity).  In this case, the widening of the late L sequence is caused primarily by the turnover in
colors at the onset of the transition.  While the observational evidence is consistent with a roughly flat SFR over the age of
of the galaxy \citep{allen05} we consider an exponentially decreasing SFR with a characteristic 
time scale of $\tau=5\,$Gyr \citep{ms79}. This results in a larger fraction of old BDs and a very narrow distribution
in the CMD (Fig. \ref{fig:panels_variations}c) that bears little resemblance with the data. In panel {\it d}, we
show the effect of using the log-normal IMF of \citet{cbah05} which greatly reduces the number of low mass objects
compared to our fiducial power law IMF with $\alpha=1$. This reduces the number of low $\teff$ and low-gravity objects 
and the sequence becomes somewhat narrower for $M_K \gtrsim 13$.

The importance of binaries in brown dwarf population synthesis
has been pointed out by \citet{chabrier02} and \citet{burgasser07}.
Substellar objects have been the target of several high resolution imaging surveys aimed at establishing the 
statistics and properties of binary brown dwarfs.  While the number of resolved systems remains small ($\sim 30$),
the observed binary fraction is $11^{+6}_{-3}$\%. Estimating the contribution of unresolved systems,
the binary fraction is somehwere in the range $\epsilon_b \sim 0.1-0.6$ \citep{burgasser07}.
The mass ratio $q=M_2/M_1$ distribution strongly favors equal mass binaries and can be approximated by 
$f(q) \sim q^4$ (\citet{burgasser07} and references therein).  
For the purpose of illustrating the role of binaries in the synthetic CMD we take a fraction
$\epsilon_b=0.3$ of the simulated single star population and assign each of them a companion of the same age and 
with a mass drawn from the above mass ratio distribution. The magnitudes of the components are combined to get the 
magnitudes of the (assumed) unresolved binary.  The result is shown in Fig. \ref{fig:panels_variations}e, where
unresolved binaries are shown in light blue and single dwarfs in green.  Because the mass ratio distribution
is sharply peaked at $q=1$, the binaries systems form a sequence that is approximately that of single dwarfs shifted
upward by 0.75 magnitude.  The sharp boundaries of the sequence (on the blue side for $M_K<12$ and on the faint side
of the transition region) remain well defined even though the total dispersion has increased significantly. Unless the
binary mass fraction is well under our choice of $\epsilon_b=0.3$ and the mass ratio distribution is much flatter
than $q^4$, unresolved binaries contribute significantly to the breadth of the dwarf sequence at all $\teff$.

Finally, we consider the effects of variations in metallicity on the CMD. The large grids of cloudy atmosphere
models with non-solar metallicity required for such a calculation are not yet available. We can presently 
explore the effect of metallicity on cloudless atmosphere models with [M/H]=-0.3, 0 and +0.3. 
These are appropriate for T dwarfs beyond the L/T transition.  
Separate evolution sequences were computed for each of those three values of [M/H]. 
In addition to the input mass and age distributions, the synthetic population is drawn from the observed distribution in 
metallicity
of 253 M dwarfs \citep{casagrande08} spanning $-0.9 \le {\rm [M/H]} \le 0.3$ with and average of $-0.20 \pm 0.22$. 
Our model grids encompasses 73\% of the
stars in this distribution so extrapolations of the evolution and colors to metallicities below $-0.3$ play a small role. For this calculation,
we adopt a power law IMF with $\alpha=1$ and a constant SFR for ages of 0--10$\,\rm Gyr$, as in Fig. \ref{fig:panels_fsed}e.
As we found for cloudless models with solar metallicity, the distribution is systematically too blue in $J-K$ by about 0.4 magnitudes,
but the slope and width of the distribution is quite similar to that of the observed late T dwarf sequence. Thus metallicity
is likely to be a significant contributor to the ``second parameter'' for the late T dwarfs.

\subsection{Color-Magnitude diagrams for clusters}

Brown dwarfs in galactic clusters offer the advantages of a uniform population
with a known distance, metallicity and a very narrow age distribution compared to field
objects. On the other hand, brown dwarfs in clusters are typically much more distant
than typical field brown dwarfs and there are very few cluster L and T dwarfs known
because they are extremely faint. They could also be rare if the IMF decreases towards lower
masses, as with the log-normal distribution of \citet{chabrier02}.  Technological progress will undoubtedly help
mitigate these limitations and we anticipate the emergence of new diagnostics of the evolution of brown
dwarfs from the study of galactic clusters.
We have generated synthetic populations of 1000 objects with masses between 0.006 and 0.1$\,M_\odot$ and an 
IMF index of $\alpha=0.6$ and for ages of 10, 50, 100, 200 and 500$\,$Myr, with our hybrid evolution sequence.  
This value of $\alpha$ is typical for galactic clusters \citet{moraux03,bouvier03,byn02,byn04}, although a 
recent study suggests that the IMF of young clusters is decreasing in the substellar regime \citep{andersen08}. 
The SFR is assumed constant for $\pm 1\,$Myr around
the age of each cluster, except for the 10$\,$Myr case where we used  a dispersion of $\pm 0.5\,$Myr. The resulting ($\teff$, $\log g$) distributions,
which follow isochrones of the hybrid evolution sequence, are shown in Fig. \ref{fig:cluster_hybrid}. 
 The
phase of deuterium burning is particularly prominent at these relatively young ages and occurs within the
modeled transition for ages of 50--200$\,$Myr. Inflexions in the isochrones at the
$\teff$ end points of the modeled L/T transition are also apparent.

All five synthetic clusters are shown in the $M_K$ vs $J-K$ CMD (Fig. \ref{fig:CMD_clusters}), where
objects with $\teff > 2400\,$K have been removed as they fall outside of our surface boundary condition
in a regime where extrapolation is unreliable.  This figure is to be compared with the CMD for the field
population (Fig. \ref{fig:CMD_disk_hybrid}).  In relatively young clusters (10 -- 50$\,$Myr), the L/T
transition region occurs at 1.5 to 2 magnitudes brighter in $M_K$ than for field brown dwarfs, on
account of their lower gravity.  The $J-K$ of the reddest brown dwarfs in a cluster increases steadily
with the cluster age, under our assumption that the $\teff$ of the onset of the transition is independent of
gravity.  

The most unusual (and to our knowledge, not previously predicted) feature of the cluster CMD is a backtracking loop along each isochrone
except the oldest (500$\,$Myr).  It is not plotted on the youngest sequence (10$\,$Myr)
because it occurs at $\teff \sim 2600\,$K, above the limit of our evolution tracks.  This feature
arises because at ages greater than about $10^6\,\rm yr$, brown dwarfs in the deuterium burning phase 
(which occurs at lower $\teff$ and larger $M_K$ for older 
clusters (Fig. \ref{fig:cluster_hybrid})) are more luminous  than {\it both} less massive and somewhat
more massive objects. It is a simple consequence of the multi-valued gravity (i.e. luminosity)
for a given $\teff$ (i.e. color) around the phase of deuterium burning (Fig. \ref{fig:cluster_hybrid}).
This feature should be most distinctive and detectable in clusters with ages of the order of
$\sim 50$ to 100$\,$Myr.   This feature is also visible as a bump in the cluster LF, as seen in the simulations of
\citet{allen03}.

\subsubsection{Pleiades}

We compare our simple model of the L/T transition with the results of deep brown dwarf surveys in the
Pleiades cluster.  The relevant parameters of the cluster are its  distance ($134\pm3\,$pc, \citet{percival05}), 
age ($\sim 100-120\,$Myr, \citet{martin98}), metallicity ([Fe/H]$=-0.034\pm0.024$, \citet{bf90}), and extinction
($E(V-I)=0.06$, \citet{sh87}).  Deep surveys have established that its low mass stellar and substellar population 
can be reproduced by a power law IMF with an index of $\alpha \sim 0.6$ \citep{moraux03,bihain06}.
We generate a synthetic brown dwarf population with our hybrid evolution sequence ([M/H]$=0$),
with $\alpha=0.6$ and stellar ages uniformly distributed between 105 and 115$\,$Myr.\footnote{This probably overestimates the age spread
of the cluster but this hardly matters here, as shown by the narrow sequence of the synthesized population in 
Fig. \ref{fig:CMD_pleiades_bihain}.}  The synthetic
cluster sequence is compared with the data of \citet{bihain06} and \citet{casewell07} in Figs. \ref{fig:CMD_pleiades_bihain} and
\ref{fig:CMD_pleiades_casewell}, respectively, where we assume that $E(J-K)=A_K=0$ and $d=134\,$pc. 
The \citet{casewell07} survey goes $\gtrsim 1$ magnitude fainter and shows a tighter L dwarf sequence.
Both figures show that the modeled L dwarf sequence is too blue by $\sim 0.3$ magnitude, too faint by $\sim 0.5$ magnitude, 
or a combination of both effects.  The
\citet{casewell07} data (Fig. \ref{fig:CMD_pleiades_casewell}) goes deep enough to reveal two objects that appear to be early T dwarfs
in L/T transition for the cluster, but they
are a good magnitude fainter than the modeled population.  Given that this hybrid
evolution/color model agrees fairly well with the local field population of brown dwarfs (Fig. \ref{fig:CMD_disk_hybrid}),
the discrepancy shown in Figs. \ref{fig:CMD_pleiades_bihain} and \ref{fig:CMD_pleiades_casewell} is somewhat surprising.

The part of the modeled sequence that
corresponds to the M and L spectral types ($M_K<12$) is $\sim 0.4\,$ magnitude too blue in $J-K$. A younger age for the sequence 
would improve the agreement but a cluster age of $30-40\,$Myr is required to make it overlap the data --
an implausible explanation.  Figure {\ref{fig:all_data} shows that this departure from the data is not much worse than we
found for the disk population (\S4.1) and in all probability is caused by the same model deficiency.
The two dimmest objects that appear to be in the L/T transition region \citep{casewell07} are almost one magnitude
fainter than predicted by our simple transition model.
These T dwarf candidates could be matched by lowering the $\teff$ regime
of the transition, which is set to $1200-1400\,$K in our hybrid sequence to match the CMD of field brown dwarfs. These
two Pleiads suggest the possibility that the onset of
the L/T transition occurs at lower $\teff$ in lower gravity objects.  This hypothesis could be tested with a detailed study
of a larger sample of dim Pleiades brown dwarfs in the range of $K\sim 17-19$, for example.  It is supported by the
low $\teff$ found for two transition brown dwarf companions of young main sequence stars, HD 203030 B \citep{mh06} and
HN Peg B \citep{luhman07}.
This may be an important clue as to the physical trigger of the ``cloud collapse'' that characterizes the transition.  
On the other hand, the association of these two brown dwarfs with the Pleiades is uncertain due to
rather large error bars in their parallaxes.  Interpretation of the mismatch should be regarded with caution until membership 
can be confirmed.  Follow-up spectroscopy should unambiguously reveal their nature.
Finally, the phase of
deuterium burning causes a well-defined feature in the transition region of the near-infrared CMDs that is potentially 
observable in this cluster if more members with $K \sim 16.5$ -- 18 can be identified. In the field, this feature is
not discernible because all the primordial deuterium has been consumed in all but the very lowest mass objects for
ages above 1$\,$Gyr (Figs. \ref{fig:evo_nc} and \ref{fig:Teff_g_nc}).

Figure \ref{fig:all_data} combines both the field and Pleiades data as well as the corresponding
modeled sequences.  The Pleiades M and L sequence objects are brighter than the those in the field due do
their larger radii at a much younger age.  Above  the transition, the modeled sequences are both systematically too blue (or too faint, but that
is less likely) but show the same behavior as the data with the same approximate variation between the two populations.  In both cases, the 
maximum $J-K$ reached as well as the slightly redder value for the older field
population is well reproduced.  Interestingly, the two faintest Pleiades candidates appear to be in the L/T 
transition fall within the field dwarf transition region. 

\section{Conclusions}

Our calculation of the evolution of very low mass stars, brown dwarfs and planetary mass objects produces models that
are quantitatively in very good agreement with published calculations.  A detailed comparison
shows some systematic differences that can be attributed to conductive energy transport, different choices of composition 
for the interior,
and for the initial state but the primary source of discrepancy is the surface boundary condition.  The cloudless
and cloudy model atmospheres from which we extract the surface boundary condition have been validated with extensive comparisons 
with spectroscopic and photometric data.  Thus, our boundary condition is quite realistic and we do not expect that the
foreseeable improvements in atmosphere models will have much effect on our modeled evolution of brown dwarfs.
By using the boundary condition from our atmosphere models, we can compute self-consistently the evolution, absolute fluxes,
absolute magnitudes, and colors for a variety of cloud properties, metallicities and eventually including vertical
mixing \citep{saumon06}. Only the evolution across the L/T transition region cannot be modeled in this self-consistent
fashion because a transition cloud model is not yet available.

We have developed a simple model for the cooling and color evolution of brown dwarfs across the L/T transition. 
This hybrid model predicts an excess of brown dwarfs in the $\teff$ range of the transition by about a factor of
2 compared to purely cloudy or cloudless evolution.
We have applied this hybrid evolution model, combined with the near-infrared magnitudes predicted by our
atmosphere models, to generate synthetic CMDs that can be compared with samples of brown dwarfs in the field and in 
galactic clusters.  Our primary focus is the ``second parameter'' responsible for the dispersion about the brown dwarf
sequence in the CMD. Population synthesis is a potentially powerful tool but the results can be quite sensitive to
the input assumptions for the IMF, SFR, metallicity, and the binary frequency and mass ratio distribution.
Our knowledge of the brown dwarf population in the solar neighborhood cannot yet provide these
inputs {\it a priori}.  The relatively small number of brown dwarfs with known parallax and the 
heterogeneous nature of the sample 
imply that the observed distribution of field substellar dwarfs in the CMD is not yet fully characterized.  Nevertheless,
both observations and models have reached a stage where general trends in the ``second parameter'' along
the L-T spectral sequence can be interpreted.  

We find that for our fiducial assumptions (power law IMF with
$\alpha=1$, constant SFR over the past 10$\,$Gyr, single brown dwarfs only and [M/H]=0), the hybrid sequence 
reproduces the overall sequence from late M through late T dwarfs rather well, but not the dispersion along the
sequence.  Based on the near-infrared CMDs, we find that the L/T transition occurs between
$\teff \sim 1400-1200\,$K in field brown dwarfs, in agreement with previous estimates. While a 
transition over such a narrow range of $\teff$ appears to be ``fast'' 
considering the rather dramatic change in $JHK$ colors across the transition, these values of $\teff$
correspond to ages of 2 and 4$\,$Gyr for a 0.06$\,M_\odot$ brown dwarf. The duration of the transition 
decreases rapidly with mass however, lasting only 0.15$\,$Gyr for a 0.03$\,M_\odot$
brown dwarf.  

Better agreement can be obtained from late M to late L spectral types if the population
is younger, such as with a constant SFR that started only 5$\,$Gyr ago, by including binaries, or assuming that
there is a wider range of cloud properties for later L spectral types.  
For a fixed metallicity, all simulations predict that the distribution of brown dwarfs in the CMD will have a sharp edge
formed by old brown dwarfs of all masses ($\gtrsim 3\,$Gyr).
This edge is on the blue side of the distribution for $M_K \lesssim 12.5$ and to the red side after the $J-K$
color of the sequence turns over, corresponding to $M_K \gtrsim 13$.  This feature is not visible in the data, however, most
likely because it is blurred by variations in metallicity within the sample.  We are not able to include metallicity variations in
simulations of 
cloudy brown dwarfs, except in a very approximate way (Fig. \ref{fig:metal_f2}).  We find that it could be a significant
contributor to the second parameter.  Detailed spectral analysis of brown dwarfs with unusual $J-K$ colors for
their spectral types have more extreme cloud parameters \citep{burgasser08,cushing08,stephens08}, 
which strongly suggest that cloudiness is the second parameter. Spectral analysis with models of non-solar metallicities
have barely begun, however, so it would be premature to attribute the dispersion along the L sequence entirely
to cloud characteristics.
On the other hand, a simulation of cloudless models with an empirical metallicity distribution shows a good match
to the dispersion of the late T dwarfs. For those coolest dwarfs, we find that the second parameter is a combination
of metallicity variations (which dominate) and binaries.  Gravity is not important as the with of the distribution
is not affected by the choice of SFR or IMF.
To summarize, we find that there is no single second parameter that accounts for the dispersion of brown
dwarfs around the sequence seen in the CMD. A young age distribution, a range of metallicities and cloud properties as
well as binaries all contribute to the dispersion.  The challenge will be to untangle their contributions.

The hybrid model fares somewhat worse when compared to the much younger brown dwarf population of the Pleiades.
If the two faintest Pleiads reported are indeed T dwarf members of the cluster, then they provide strong evidence
that the L/T transition occurs at lower $\teff$ in lower gravity objects (i.e. younger or less massive).
Finally, isochrones in CMDs clearly reveal the phase of
deuterium burning at young ages, a feature that should be observable in young clusters with ages between of $\sim 50-100\,$Myr. 

At this time, the modest size of the sample of L and T dwarfs with known parallax and the lingering problems in 
modeling the atmospheres of cloudless and cloudy brown dwarfs restrict how much we can learn from the study
of CMDs.  Model limitations will eventually be overcome as new moelcular line lists are being
developed for key molecules and cloud models become more sophisticated. We anticipate a rich harvest of brown
dwarf parallaxes from the volume limited solar neighborhood census component of the Panoramic Survey Telescope 
\& Rapid Response System
(Pan-STARRS) \footnote{\tt http://pan-starrs.ifa.hawaii.edu/project/reviews/PreCoDR/documents/scienceproposals/sol.pdf}.
Color-magnitude diagrams are a potentially powerful tool for the study of brown dwarf evolution and of
the L/T transition. Statistical comparisons with synthetic populations in two-dimensional parameter space
will become an important complement to the detailed studies of the spectra of individual 
transition objects and of brown dwarf binaries.

\acknowledgments

We acknowledge support from NASA grants NAG 2-6007 and NAG 5-8919 (M. S. M.).  Support for this work, part of the Spitzer Space 
Telescope Theoretical Research Program, was provided by NASA.  We thank A. J. Burgasser for thoughtful comments
on the manuscript and J. R. Stauffer for useful discussions.

\appendix
\section{Typographical errors in Saumon, Chabrier \& Van Horn (1995)}

Typographical errors have been found in the expressions for
the thermodynamics of hydrogen and helium mixtures given in Saumon, Chabrier
\& Van Horn (1995, hereafter SCVH). The correct expressions are:

\begin{equation}
\eqnum{45}
   S_T=(1-Y){S^{\rm H} \over S}\, S_T^{\rm H} + Y{S^{\rm He} \over S} \,S_T^{\rm He}
       + {S_{\rm mix} \over S}{\partial \log S_{\rm mix} \over \partial \log T}\bigg|_P,
\end{equation}

\begin{equation}
\eqnum{46}
   S_P=(1-Y){S^{\rm H} \over S}\,S_P^{\rm H} + Y{S^{\rm He} \over S} \,S_P^{\rm He}
       + {S_{\rm mix} \over S}{\partial \log S_{\rm mix} \over \partial \log P}\bigg|_T.
\end{equation}

\begin{equation}
\eqnum{56}
   \delta={2(2-2X_{\rm He}-X_{{\rm He}^+}) \over 3(1-X_{{\rm H}_2}-X_{\rm H})} \beta\gamma,
\end{equation}
where the equation numbers are those of SCVH.

The last correction affects the calculation of the contribution of the electrons
to the ideal entropy of mixing.  In practice, this
matters only when the gas is nearly fully ionized, where the ideal entropy of
mixing $S_{\rm mix}$ (Eq. 53) is only a rough approximation of the actual entropy of mixing,
except in the high temperature, low density limit where it is nearly exact.  This correction should
not be of much concern to applications of the mixed H/He EOS to stellar and
planetary interiors where only a very small fraction of the mass is not fully ionized.

\bibliographystyle{apj}
\bibliography{references}

\clearpage

\begin{figure}
   \plotone{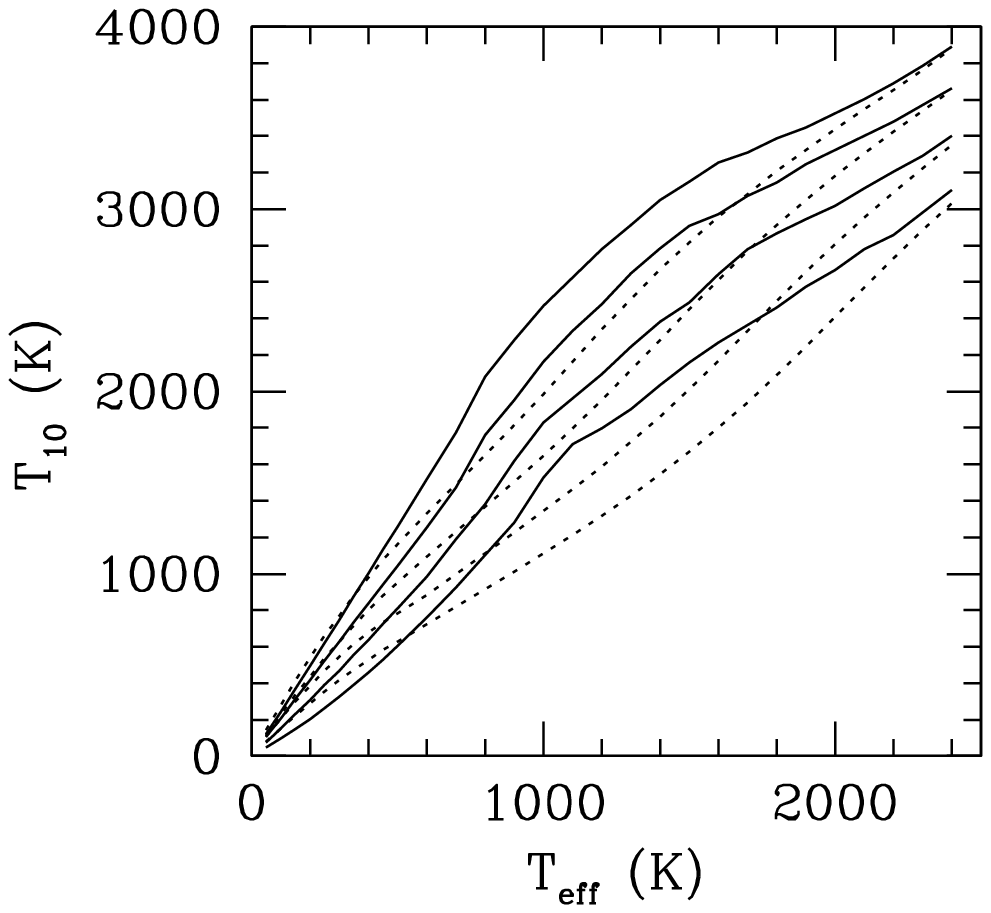}
   \caption{Surface boundary condition, expressed as $T_{10}(\teff,g)$ for solar metallicity atmospheres without clouds (dotted curves)
            and with clouds, using a sedimentation parameter $f_{\rm sed}=2$  (solid curves).  Each curves correspond to a different
            surface gravity with $\log g=4$, 4.5, 5, and 5.5, from top to bottom, respectively.  Larger values of $T_{10}$ imply a higher entropy for
            the interior. The surface boundary condition is interpolated below $\teff=500\,$K to $(0,0)$.}
    \label{fig:BC}
\end{figure}
\clearpage

\begin{figure}
   \plotone{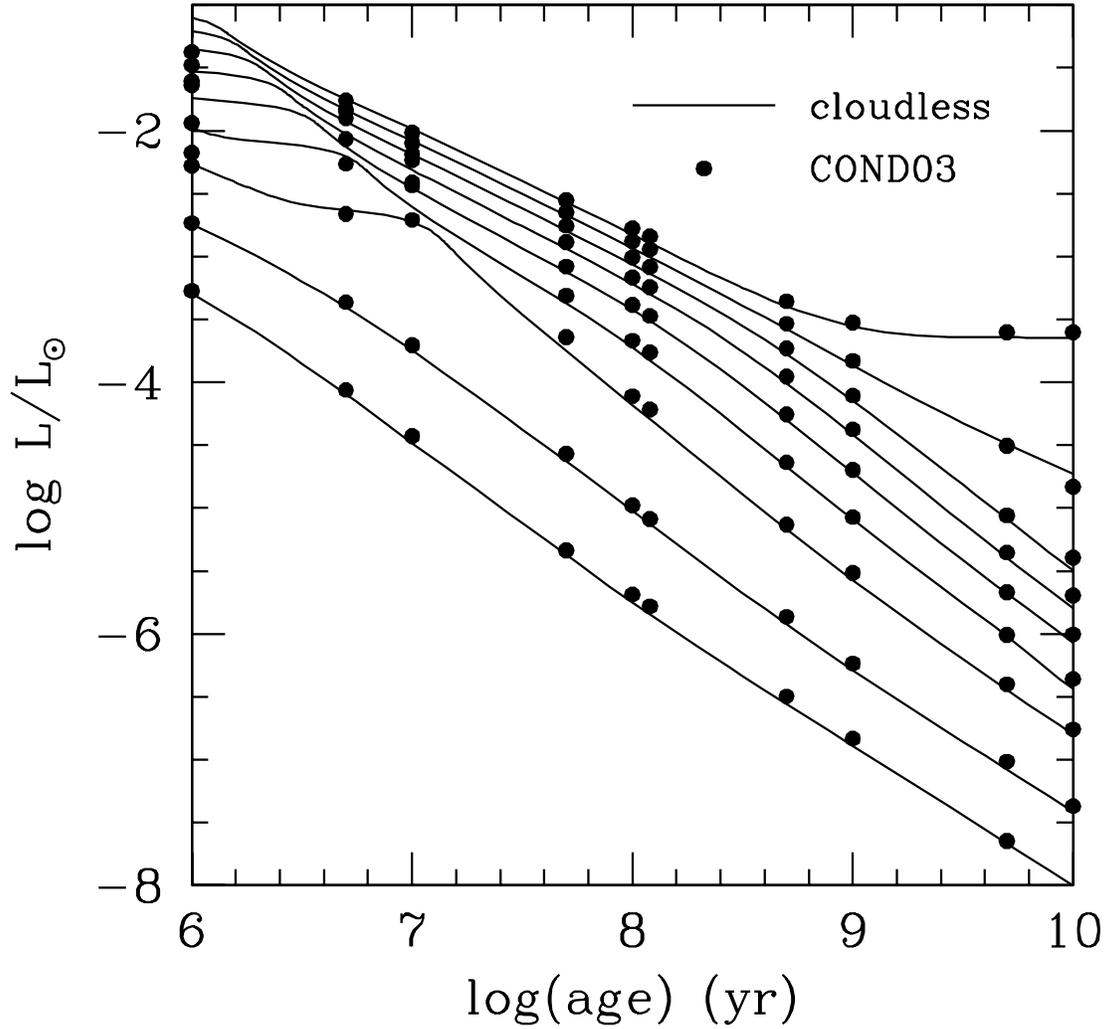}
   \caption{Evolution of the luminosity of brown dwarfs for masses of 0.005, 0.01, and up to 0.08$\,$M$_\odot$ in steps of 0.01,
            from bottom to top, respectively.  This sequence of models uses cloudless model atmospheres for the surface boundary condition.
            Solid dots show the models of the Lyon group based on the COND03 atmosphere models for the same masses (Baraffe et al. 2003).}
    \label{fig:evo_nc}
\end{figure}
\clearpage

\begin{figure}
   \plotone{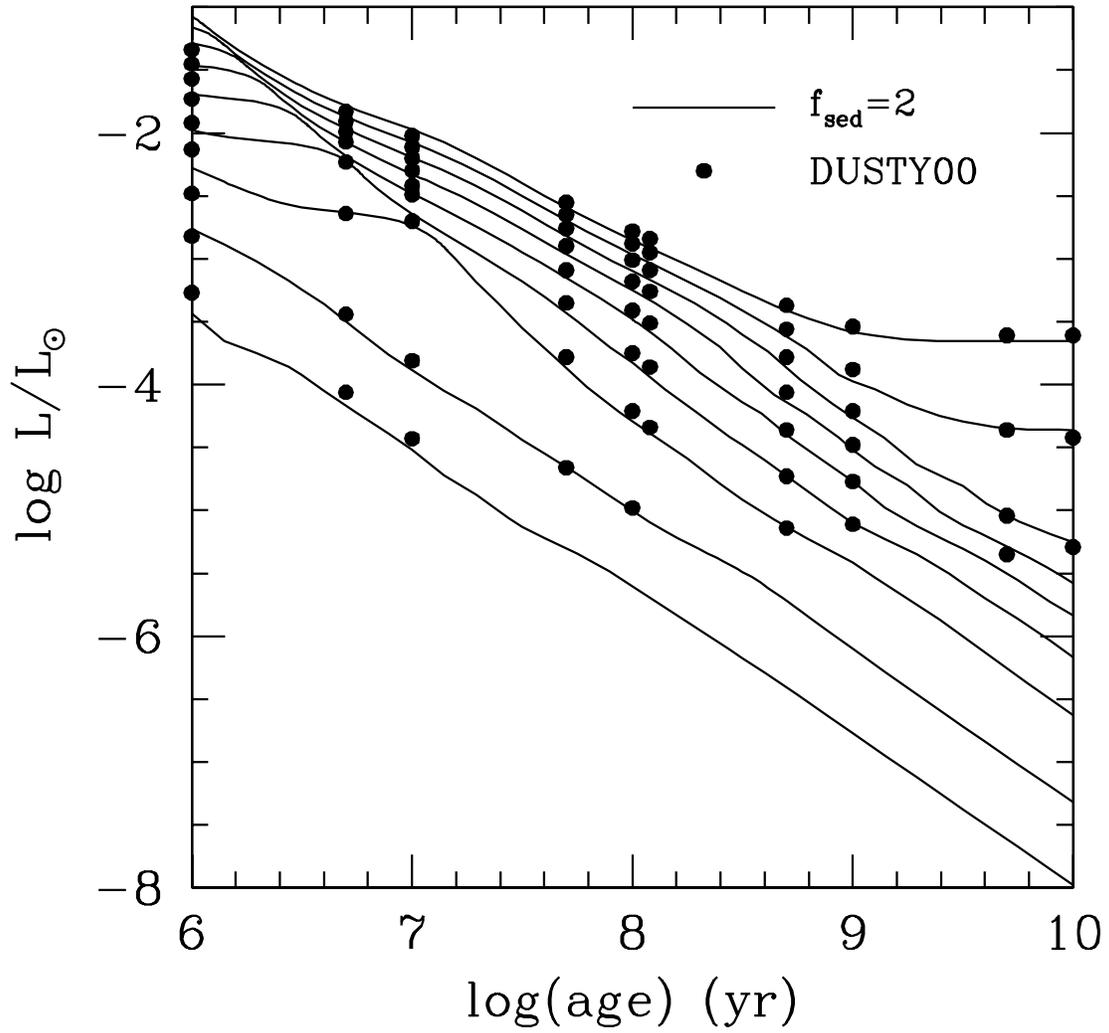}
   \caption{Same as Fig. \ref{fig:evo_nc} but with a cloudy surface boundary condition ($f_{\rm sed}=2$).
            Solid dots show the models of the Lyon group based on the DUSTY00 atmosphere models
            (Chabrier et al. 2000; Baraffe et al. 2001).}
    \label{fig:evo_f2}
\end{figure}
\clearpage

\begin{figure}
   \plotone{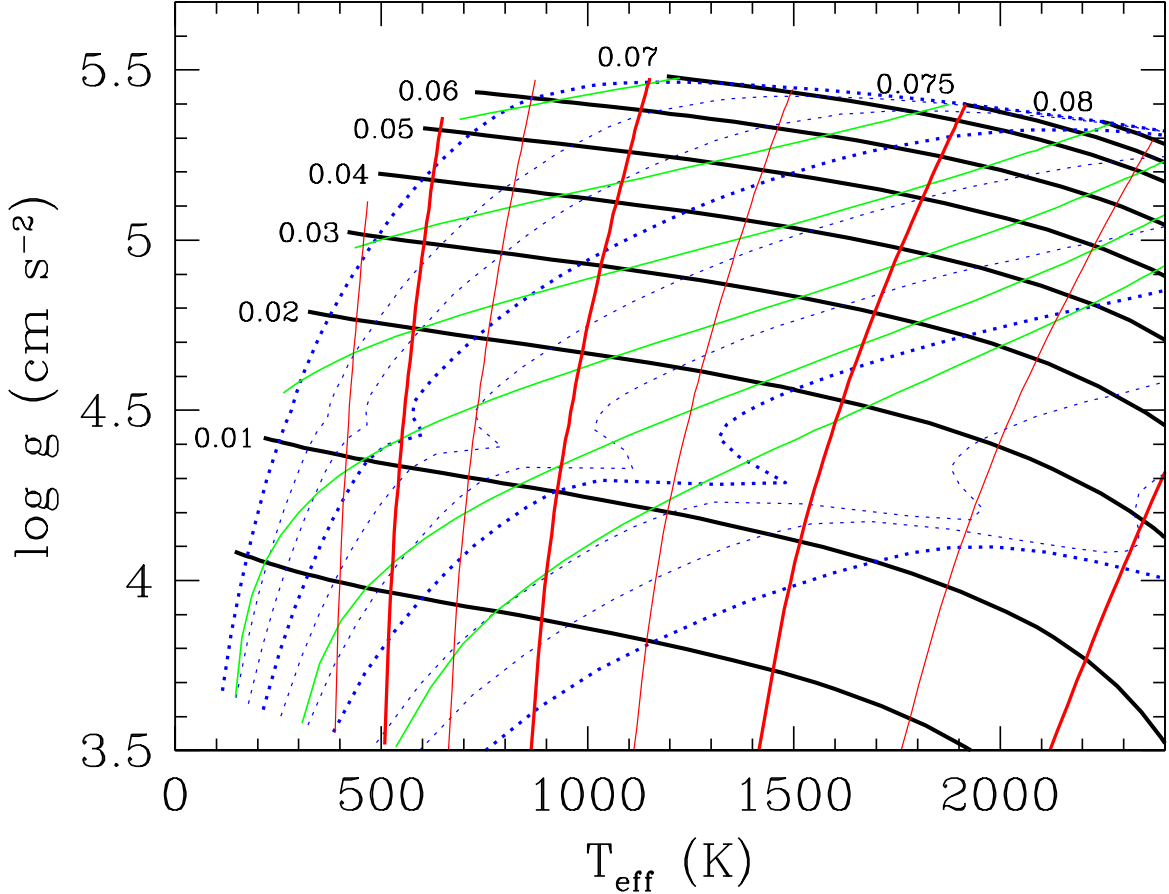}
   \caption{Evolution of brown dwarfs in $\teff$ and gravity for the sequence based on cloudless atmospheres. The evolution proceeds 
            from right to left along the heavy black lines, which are labeled with the mass in M$_\odot$. Isochrones are shown by
            the blue dotted lines: (from right to left) {\bf 0.01}, 0.02, 0.04, {\bf 0.1}, 0.2, 0.4, {\bf 1}, 2, 4, and {\bf 10}$\,$Gyr.
            The nearly vertical red lines are curves of constant luminosity: (from right to left): $\log L/L_\odot= -3$, to $-$6.5\ in
            steps of 0.5. Curves of constant radius are shown in green: (from top to bottom) 0.08 to 0.13$\,R_\odot$  in steps of 0.01.
            The phase of deuterium burning is revealed by the kink in the isochrones for objects with masses between 0.01 and 0.02$\,M_\odot$.
           [{\it See the electronic edition of the Journal for a color version of this figure.}]}
    \label{fig:Teff_g_nc}
\end{figure}
\clearpage

\begin{figure}
   \plotone{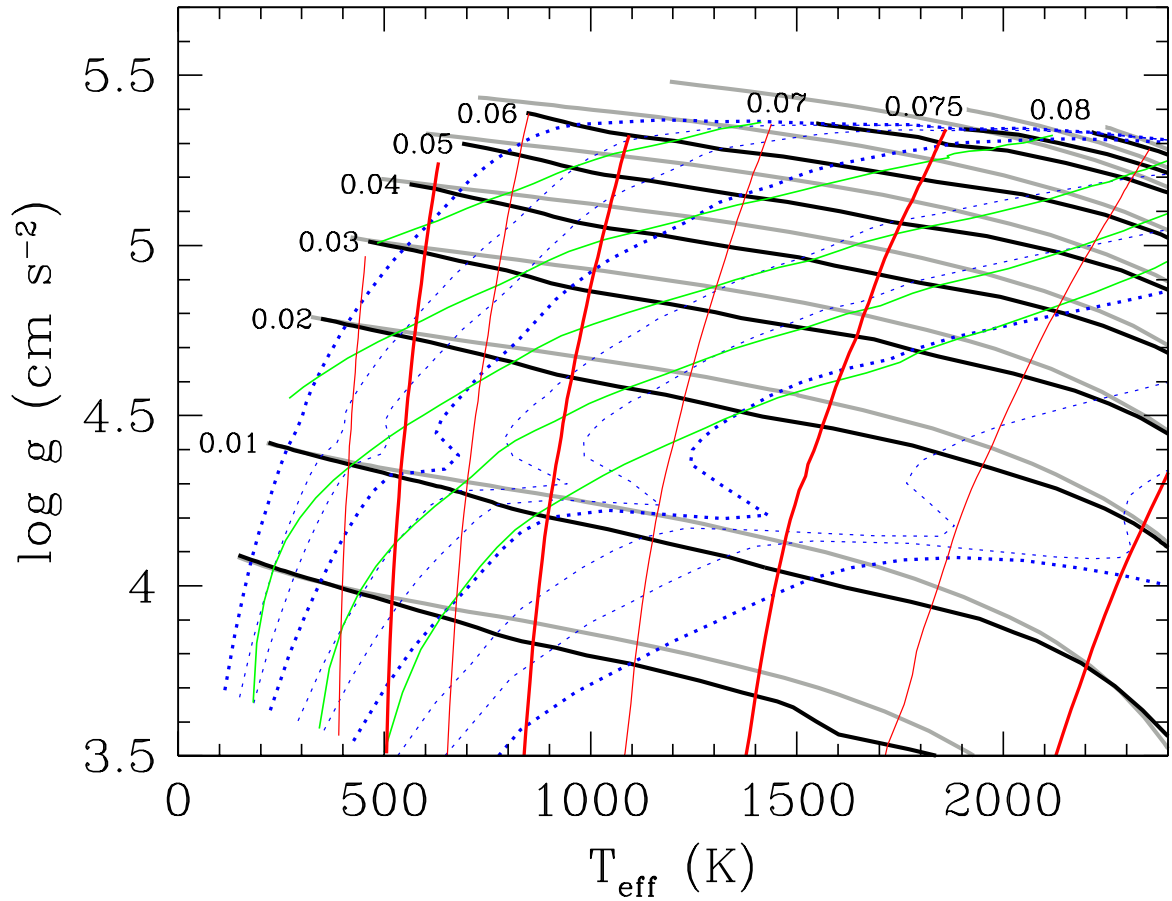}
   \caption{Same as Fig. \ref{fig:Teff_g_nc} but with cloudy atmospheres models for the surface boundary condition ($f_{\rm sed}=2$).
            The cloudless cooling tracks from Fig. \ref{fig:Teff_g_nc} are shown in gray for comparison.
           [{\it See the electronic edition of the Journal for a color version of this figure.}]}
    \label{fig:Teff_g_f2}
\end{figure}
\clearpage

\begin{figure}
   \plotone{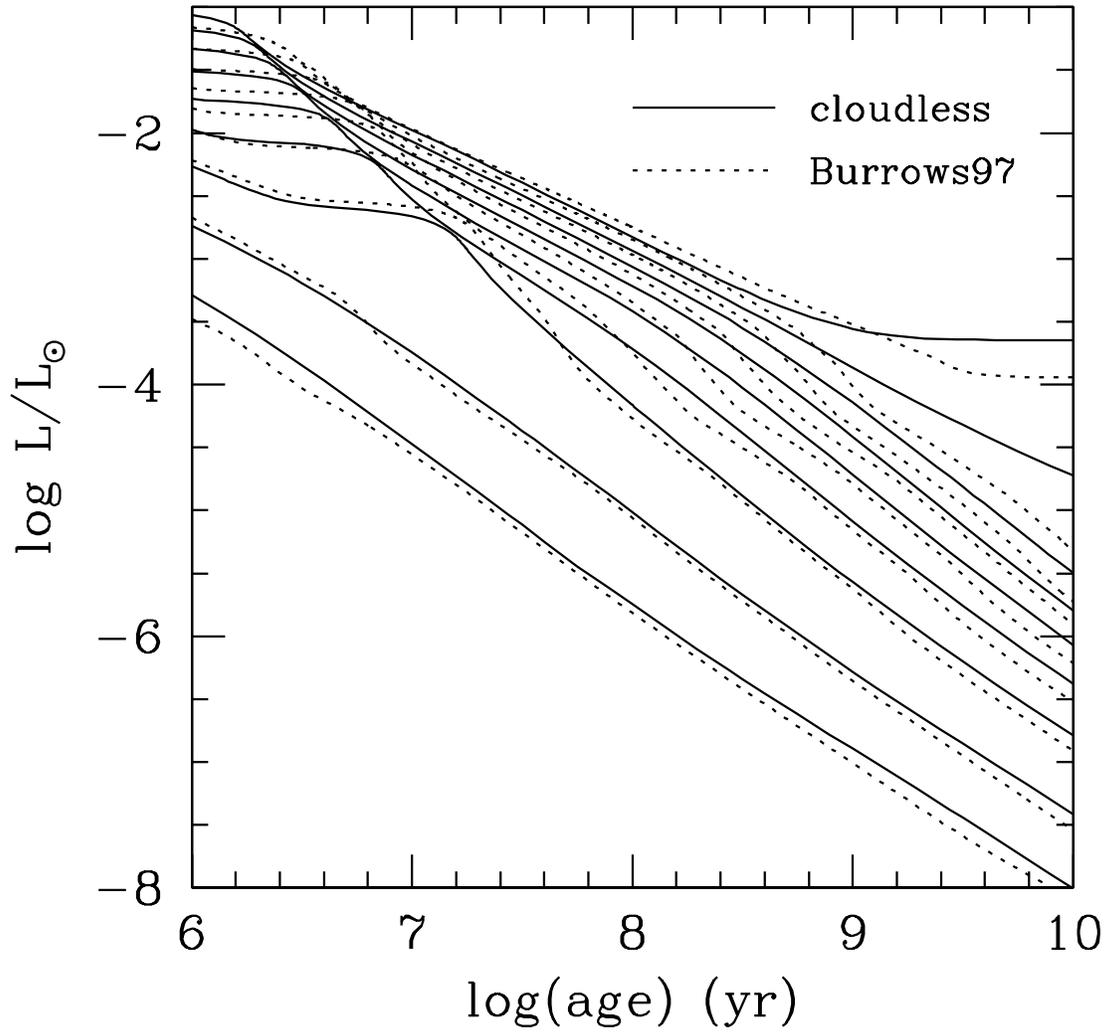}
   \caption{Evolution of the luminosity of brown dwarfs
            This shows the same cloudless sequence as in Fig. \ref{fig:evo_nc} (solid lines), compared with the
            cloudless cooling sequences of \citet{bur97} (dotted lines).}
    \label{fig:evo_B97}
\end{figure}
\clearpage

\begin{figure}
   \plotone{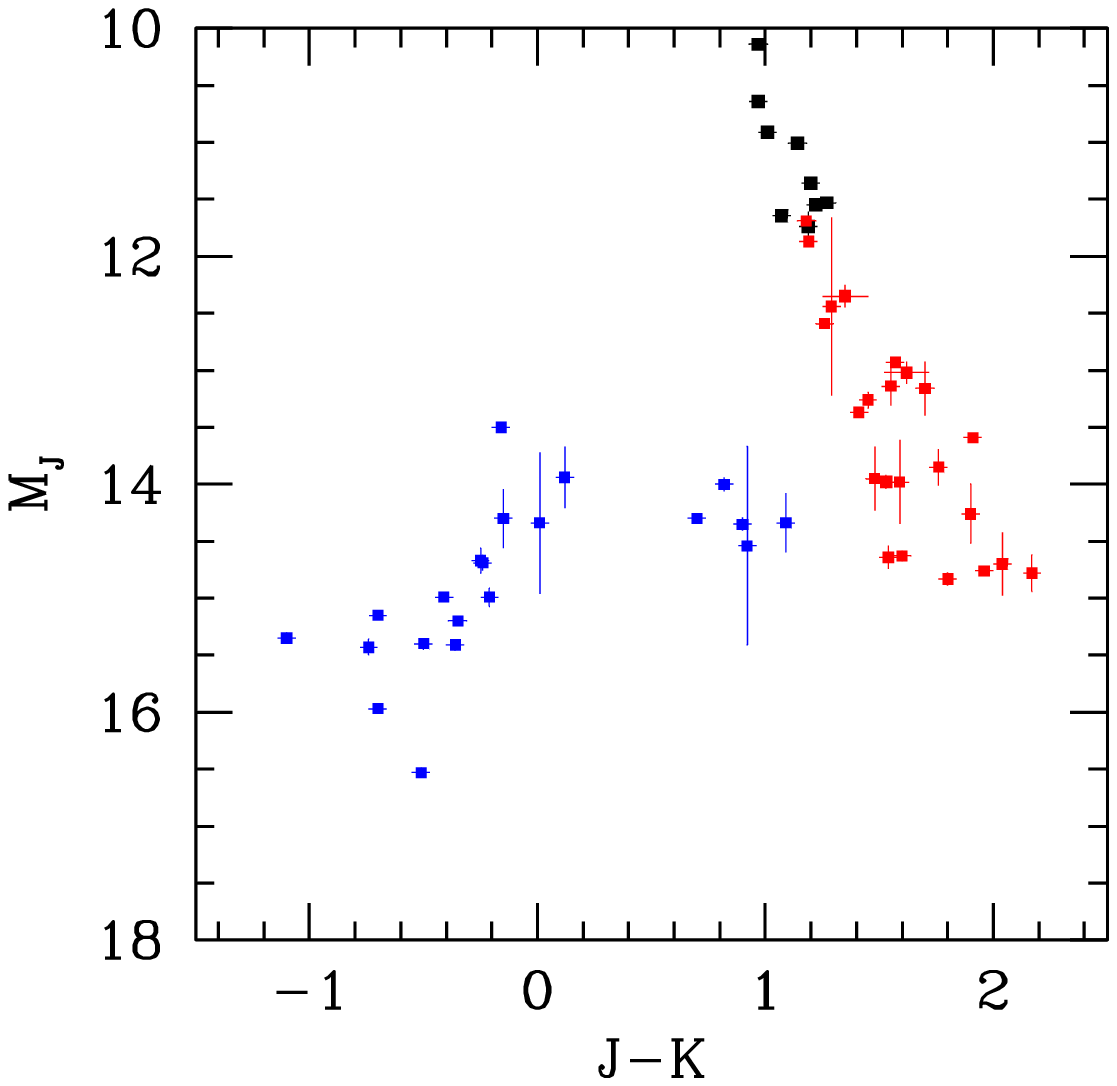}
   \caption{Color-magnitude diagram (MKO system) for field brown dwarfs, showing the photometry of
            \citet{leggett02} and \citet{knapp04} with M dwarfs in black, L dwarfs in red and T dwarfs in blue. All known binaries
            have been removed from the sample except those with resolved MKO photometry: 
            $\epsilon$ Indi B \citep{mccaugh04}, SDSS J102109.69$-$030420.1 and SDSS J042348.57$-$041403.5 \citep{burgasser06}, 
            and Kelu-1 \citep{ll05}.  The parallaxes are from \citet{perryman97}, \citet{dahn02}, \citet{tbk03}, \citet{vrba04},
            and various references in \citet{leggett02}.  
           [{\it See the electronic edition of the Journal for a color version of this figure.}]}
    \label{fig:CMD_data}
\end{figure}
\clearpage

\begin{figure}
   \plotone{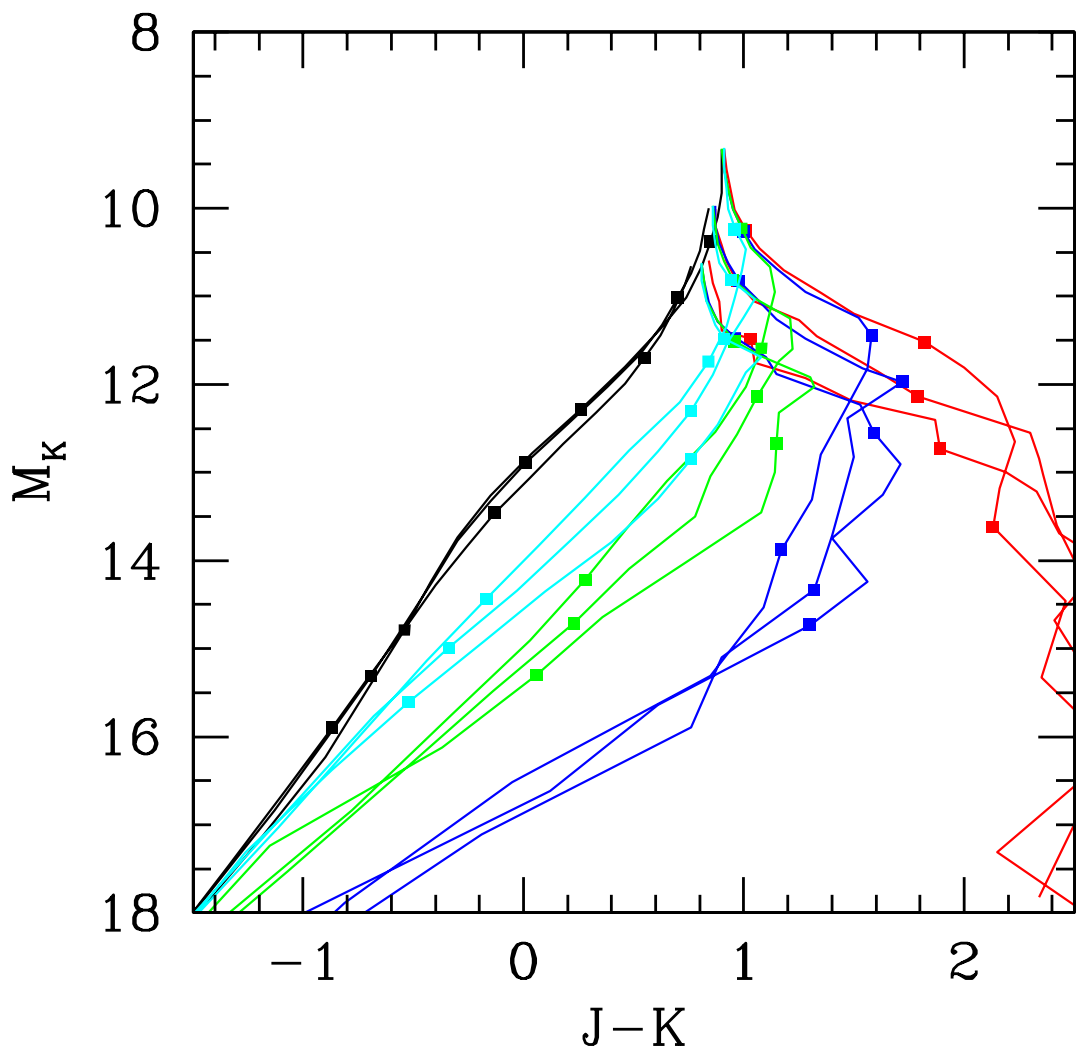}
   \caption{Color-magnitude diagram (MKO system) for atmosphere models at three gravities ($\log g=4.5$, 5 and 5.5, from right to left at
            the bright end) and for cloud condensation parameters $f_{\rm sed}=1$ (red), 2 (blue), 3 (green), 4 (cyan), and
            cloudless models (black).  The effective temperature ranges from 2400$\,$K (top) to 500$\,$K and
            squares along each curve indicate $\teff=2000$, 1500, and 1000$\,$K (from top to bottom).
            All models shown have solar metallicity.
            Some of the curves are not smooth due to numerical difficulties with a few cloudy cases.
           [{\it See the electronic edition of the Journal for a color version of this figure.}]}
    \label{fig:CMD_atmos_MK}
\end{figure}
\clearpage

\begin{figure}
   \plotone{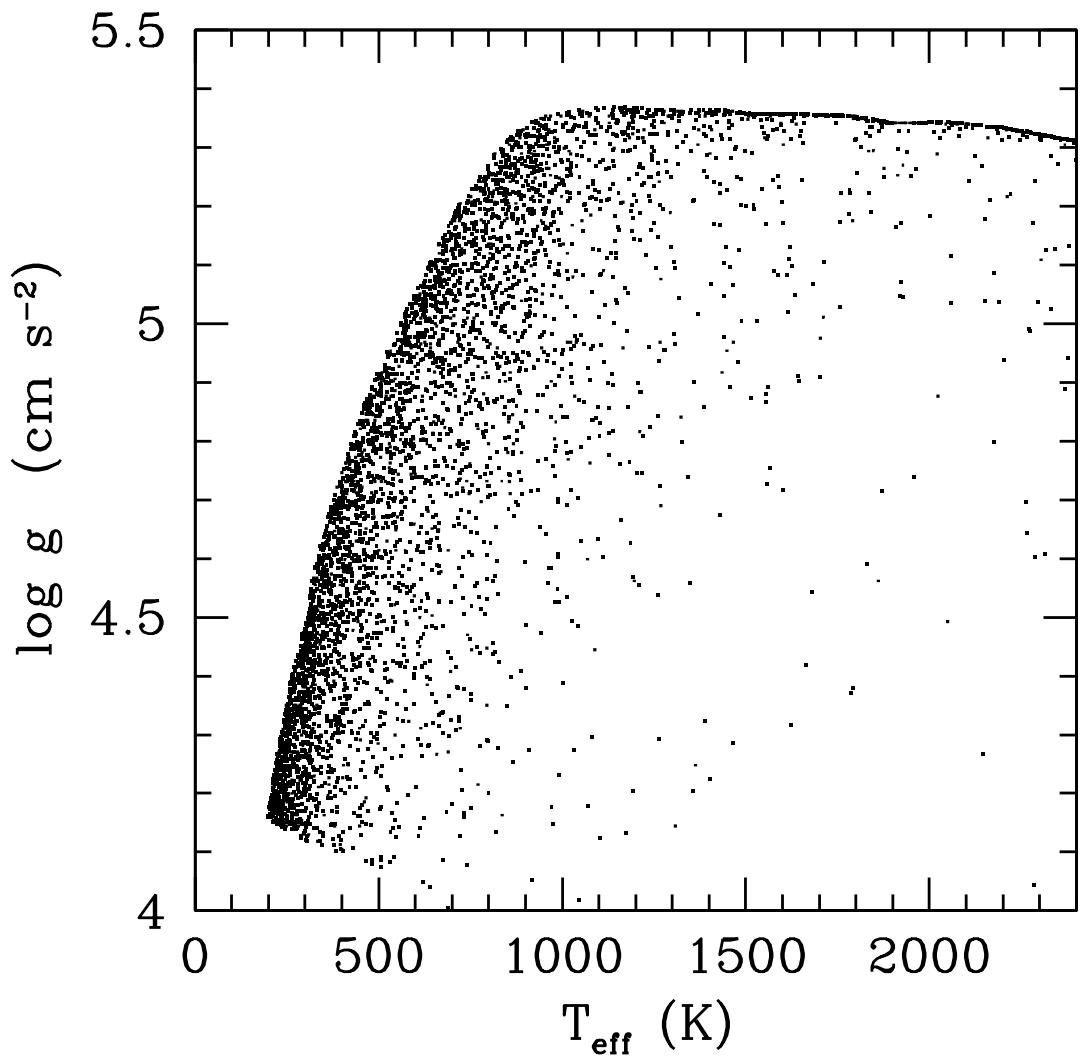}
   \caption{Distribution in ($\teff$,$\log g$) of a simulated local field population of brown dwarfs with power law IMF with index $\alpha=-1$,
           masses between 0.006 and 0.1$\,M_\odot$, and a uniform age distribution between 0 and 10$\,$Gyr.  The mapping from ($M$,age)
           to ($\teff$, $\log g$) is based on the cloudy evolution sequence with $f_{\rm sed}=2$ and [M/H]=0.}
    \label{fig:disk_distrib}
\end{figure}
\clearpage

\begin{figure}
   \plotone{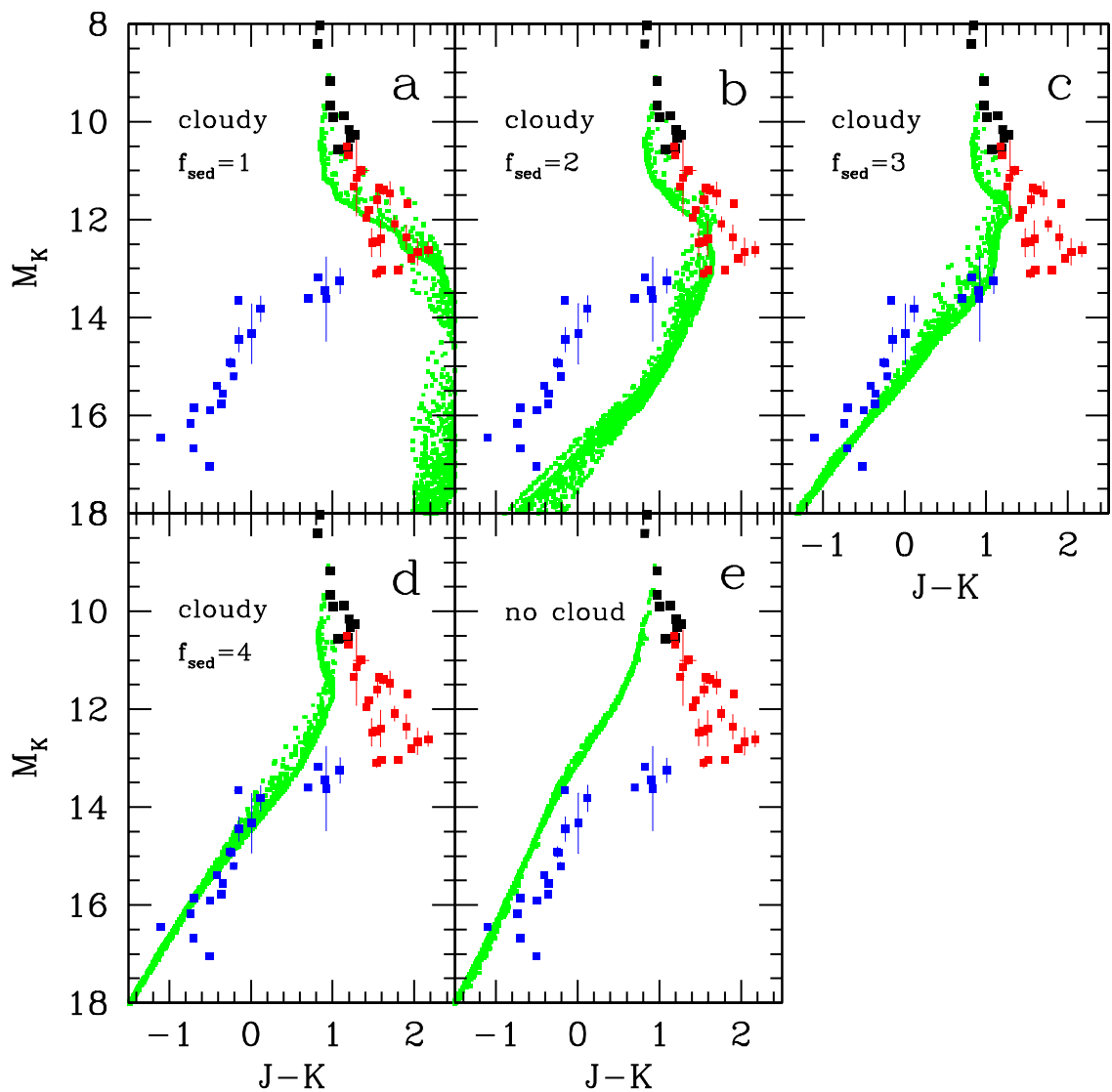}
   \caption{Color-magnitude diagram of a synthetic field brown population of solar metallicity based on the cloudy evolution sequence 
            ($f_{\rm sed}=2$) and cloudy 
            model magnitudes with $f_{\rm sed}=1-4$ (panels a-d) and for the cloudless sequence (panel e).  The synthetic population is 
            taken from Fig. \ref{fig:disk_distrib}.
            The sources of data are the same as in Fig. \ref{fig:CMD_data}.
            The small green squares show the synthetic population. All magnitudes are in the MKO system.
           [{\it See the electronic edition of the Journal for a color version of this figure.}]}
    \label{fig:panels_fsed}
\end{figure}
\clearpage

\begin{figure}
   \plotone{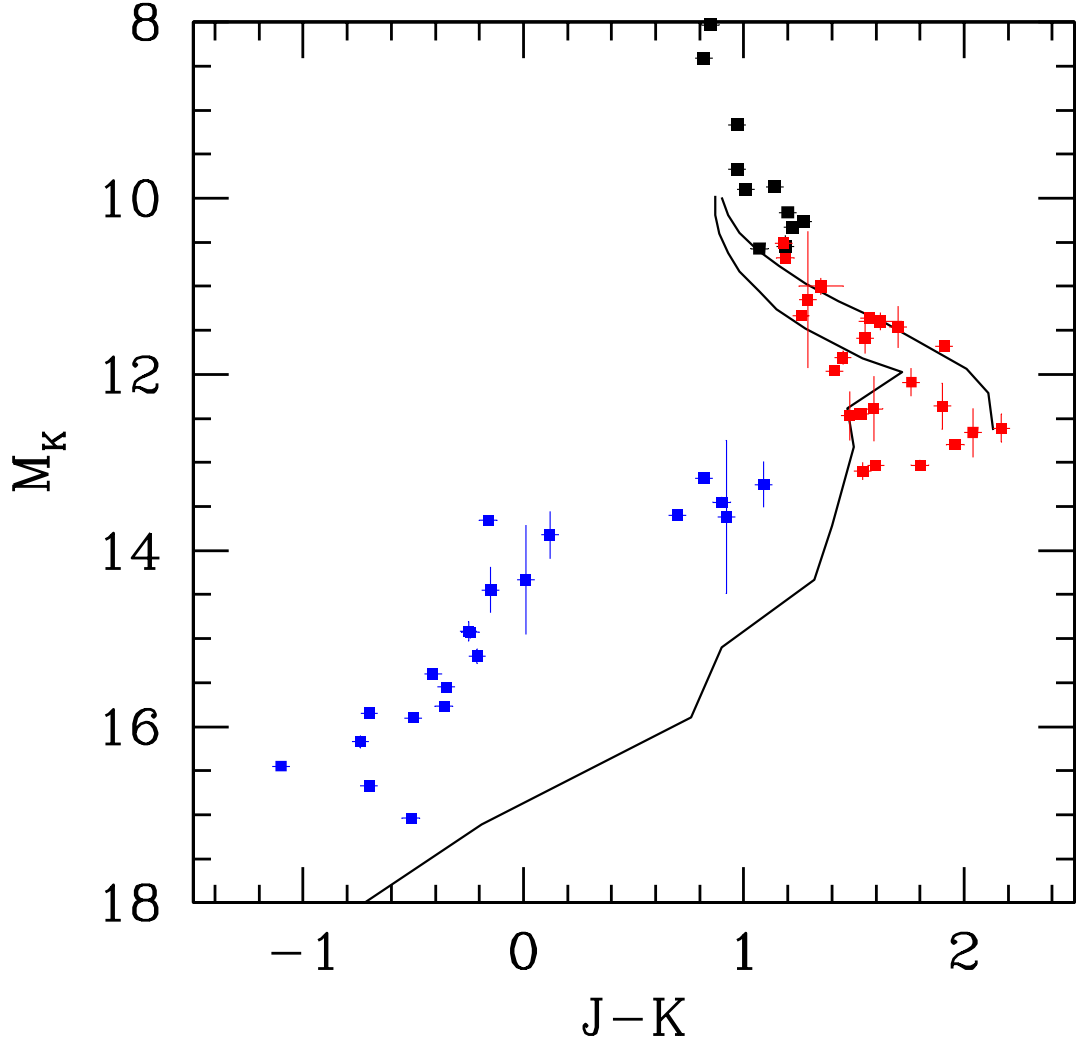}
   \caption{Colors of cloudy models with $\fsed=2$ and $\log g=5$ and two different metallicities (solid lines).  
            The curve on
            the right has [M/H]=+0.5 and extends from $\teff=2400$ to 1300$\,$K.  The curve
            on the left, which also appears on Fig. \ref{fig:CMD_atmos_MK},  has [M/H]=0 and extends from 
            $\teff=2400$ to 500$\,$K
           [{\it See the electronic edition of the Journal for a color version of this figure.}]}
    \label{fig:metal_f2}
\end{figure}
\clearpage

\begin{figure}
\epsscale{.8}
  \plotone{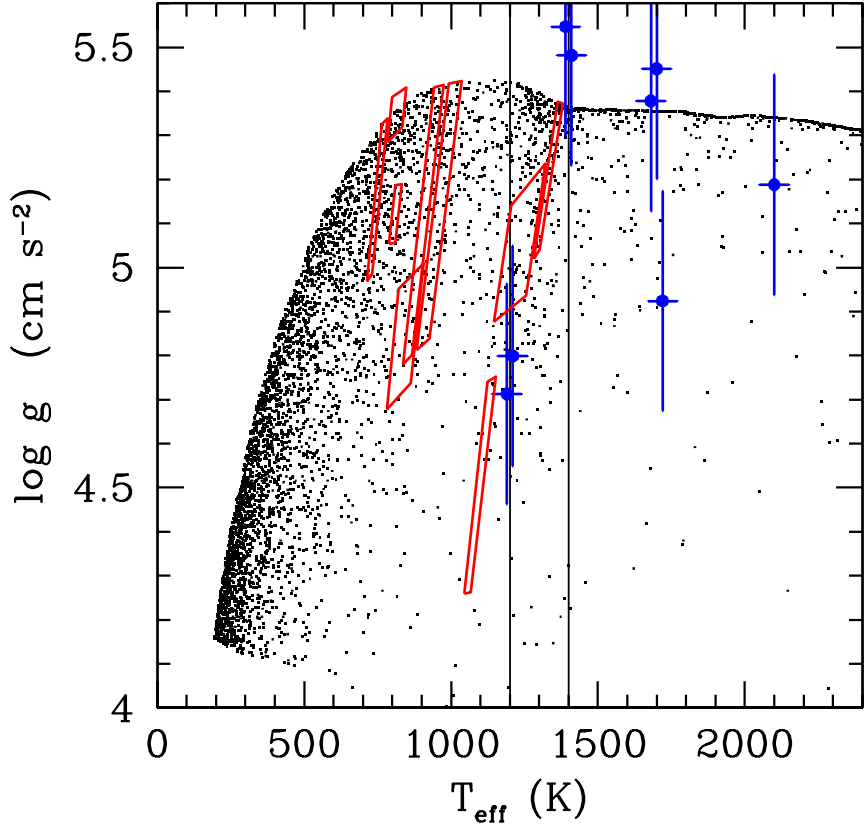}
\epsscale{1}
  \caption{Distribution in ($\teff$,$\log g$) of a simulated local field population of
          brown dwarfs with power law IMF index $\alpha=1$,
          masses between 0.006 and 0.1$\,M_\odot$, and a uniform age distribution between
          0 and 10$\,$Gyr.  The mapping from ($M$,age) to ($\teff$, $\log g$) is based
          on a hybrid cloudy/cloudless evolution sequence to mimic the L/T transition
         (see text). The vertical lines indicate the $\teff$ range of the modeled
         transition. Superimposed on the synthetic distribution are the ($\teff,\log g$)
         of field dwarfs determined by fitting their spectral energy distribution
         (\citet{cushing08}, blue).  The red boxes show the parameters of well-studied
         dwarfs that are constrained by $L$ and age.  From left to right, they are
         2MASS J0415195$-$093506 \citep{saumon07}, HD 3651B \citep{liu07}, Gl 570D
         \citep{saumon06}, $\epsilon$ Ind Bb \citep{mccaugh04},
         2MASS J12171110$-$03111131 \citep{saumon07}, 2MASS J09373487+2931409 \citep{geballe08}, Gl 229B
         \citep{saumon00}, HN Peg B \citep{leggett08}, $\epsilon$ Ind Ba
         \citep{mccaugh04}, and 2MASS J05591914$-$1404488 ($\log L/L_\odot=-4.606 \pm 0.04$ obtained with the
         method of \citet{geballe01} assuming it is a single brown dwarf).  For objects without
         age constrains available, $1-10\,$Gyr is assumed.
         [{\it See the electronic edition of the Journal for a color version of this
         figure.}]}
   \label{fig:disk_hybrid}
\end{figure}
\clearpage

\begin{figure}
   \includegraphics[angle=-90]{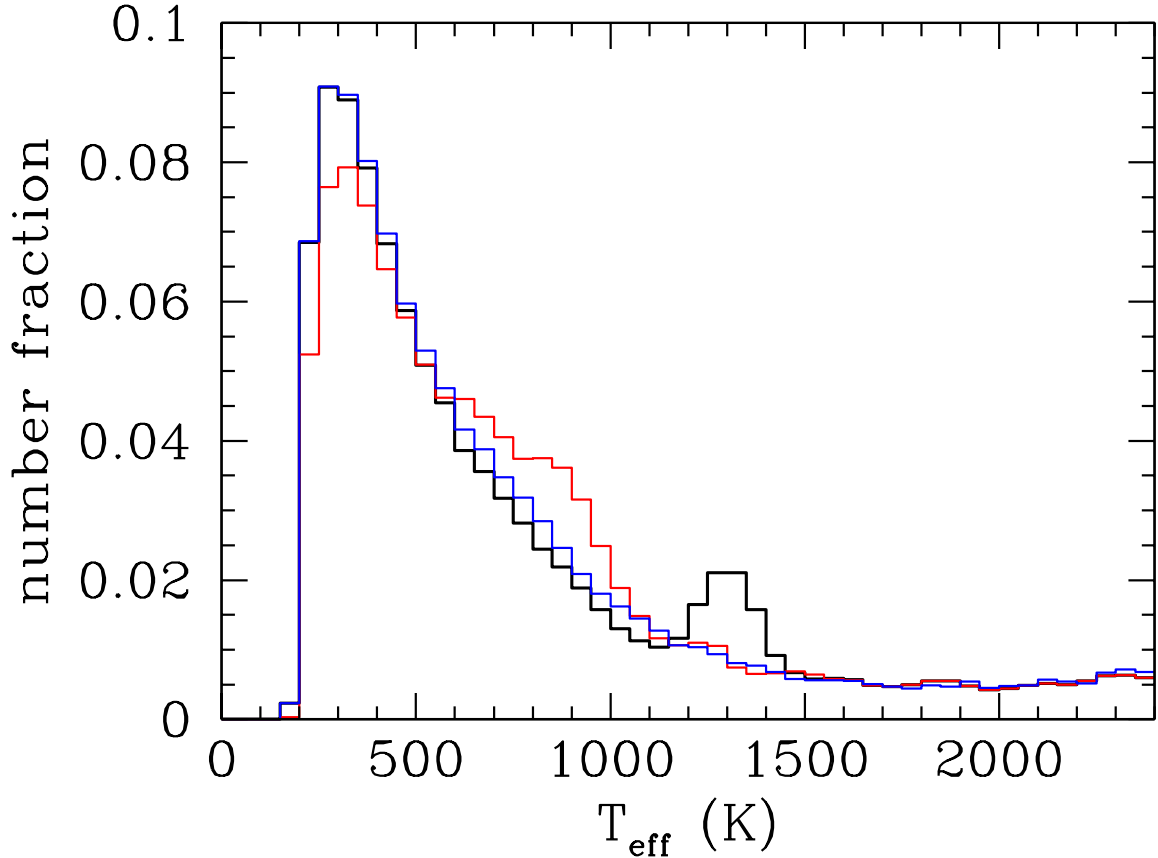}
   \caption{Distributions in $\teff$ of the synthetic disk population for the hybrid sequence (black), 
            the cloudy sequence (red) and the cloudless sequence (blue).  The distributions are
            normalized to the size of the sample. The sharp cutoff at 200$\,$K reflects the mass cutoff
            of our simulations at 0.006$\,M_\odot$.  By construction, the hybrid sequence recovers
            the cloudy sequence above $\teff=1400\,$K and the cloudless sequence below 1200$\,$K.
            In the hybrid sequence, the cooling time scale increases in the transition region (1200--1400$\,$K)
            causing a pile up of objects.  A similar effect occurs in the cloudy sequence when the cloud
            deck becomes so deep that it no longer affects the evolution.  The result is an excess of
            objects in the 600--1000$\,$K range.
           [{\it See the electronic edition of the Journal for a color version of this figure.}]}
    \label{fig:Teff_distrib}
\end{figure}
\clearpage

\begin{figure}
   \plotone{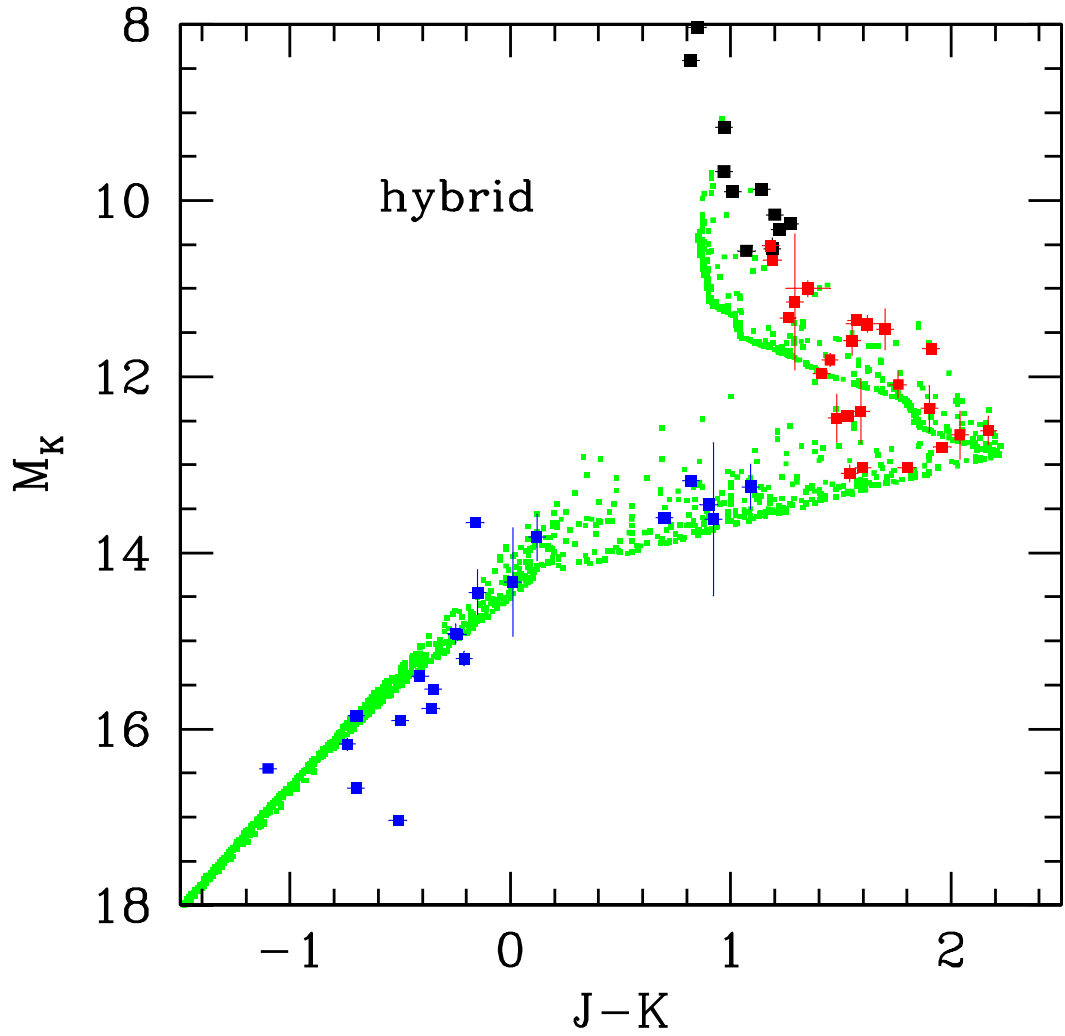}
   \caption{Synthetic color-magnitude diagram (MKO system) of a field brown dwarf population based on the hybrid evolution sequence and colors (see text).
            See Fig. \ref{fig:panels_fsed} for the legend.
           [{\it See the electronic edition of the Journal for a color version of this figure.}]}
    \label{fig:CMD_disk_hybrid}
\end{figure}
\clearpage

\begin{figure}
   \plotone{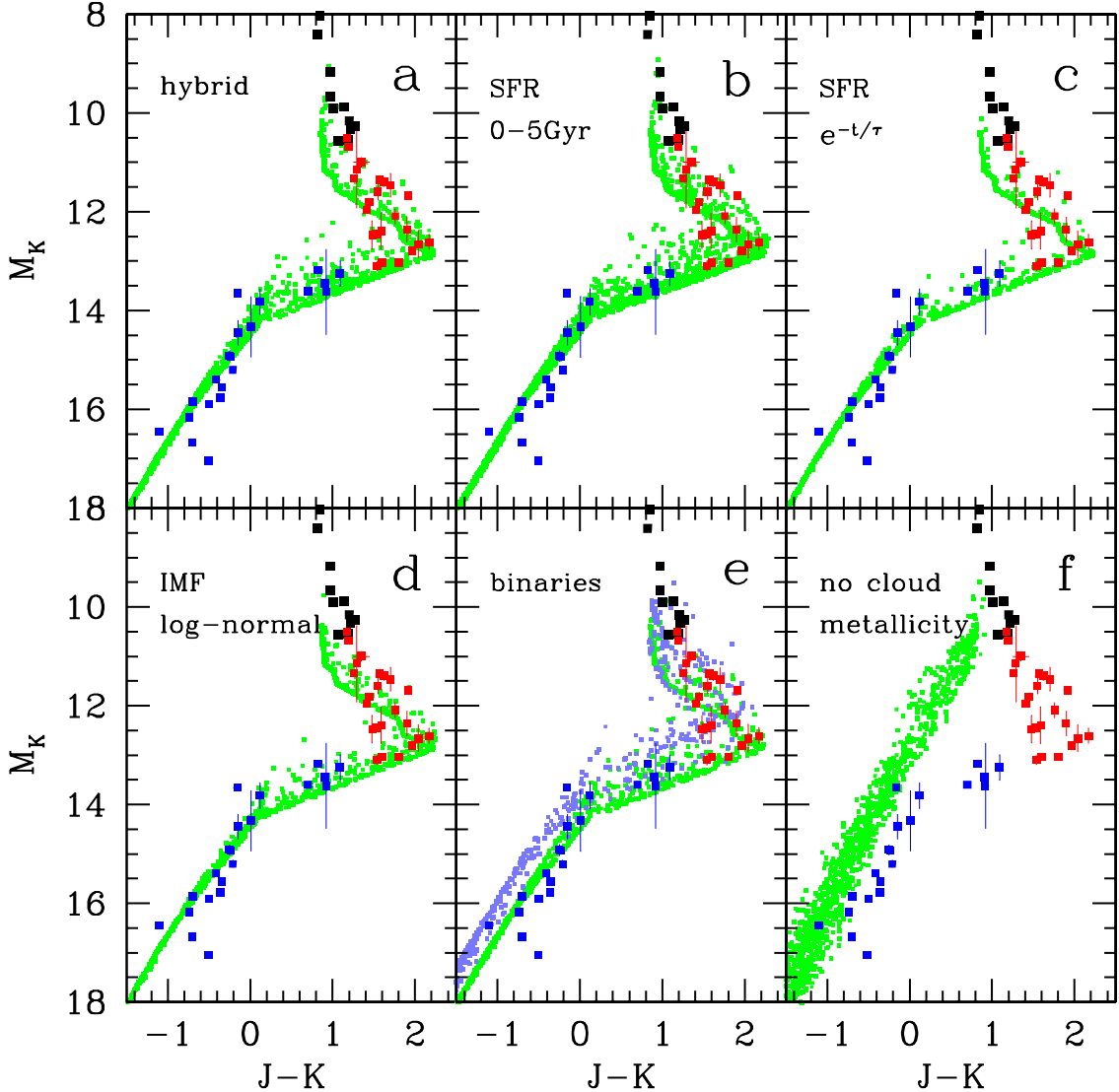}
   \caption{CMD of the simulated field population resulting from variations in the input assumptions.  
            a) fiducial sequence based on the hybrid evolution sequence with a power law IMF ($\alpha=1$),
               a flat SFR over 0--10$\,$Gyr, and single stars only (same as Fig. \ref{fig:CMD_disk_hybrid}) 
            b) same as in a) but the SFR is truncated for ages $>5\,$Gyr. 
            c) same as a) but with an exponentially decreasing SFR ($\tau=5\,$Gyr).  
            d) same as a) but with the log-normal IMF of \citet{cbah05}.  
            e) same as a) but with unresolved binaries included (light blue dots).  The binary mass
            fraction is $\epsilon_b=0.3$ and the distribution of mass ratios is $f(q)\sim q^{4}$. 
            f) cloudless models with the metallicity distribution of M dwarfs \citep{casagrande08},
              a constant SFR over 0--10$\,$Gyr and a power law IMF with $\alpha=1$.
           [{\it See the electronic edition of the Journal for a color version of this figure.}]}
    \label{fig:panels_variations}
\end{figure}
\clearpage

\begin{figure}
   \plotone{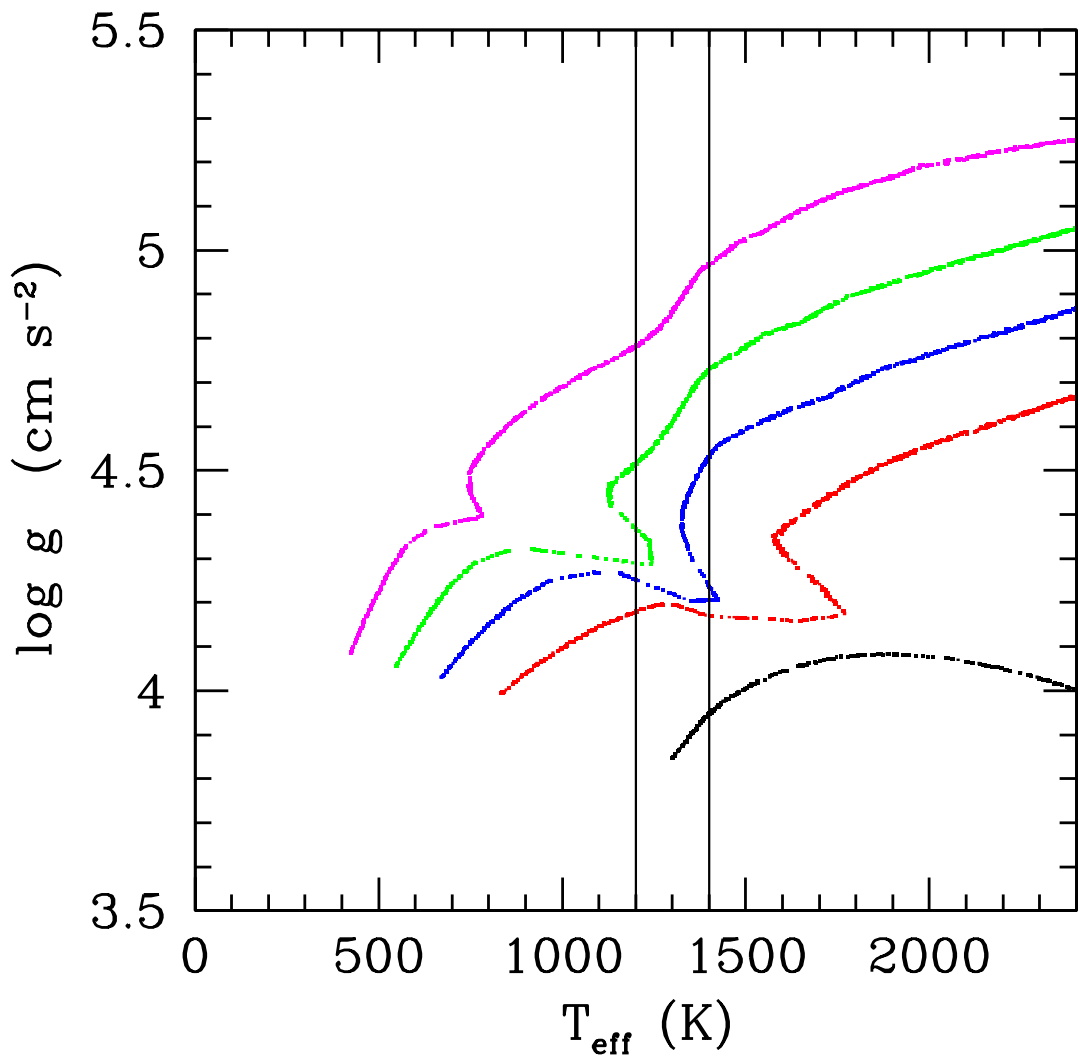}
   \caption{Distribution in ($\teff$,$\log g$) of simulated cluster populations of brown dwarfs with power law IMF index $\alpha=0.6$,
           masses between 0.006 and 0.1$\,M_\odot$, and ages of 10 (black), 50 (red), 100 (blue), 200 (green) and 500$\,$Myr (magenta).
           The mapping from ($M$,age)
           to ($\teff$, $\log g$) is based on a hybrid cloudy/cloudless evolution sequence to mimic the L/T transition.
           The vertical lines indicate the $\teff$ range of the modeled transition.
           [{\it See the electronic edition of the Journal for a color version of this figure.}]}
    \label{fig:cluster_hybrid}
\end{figure}
\clearpage

\begin{figure}
   \plotone{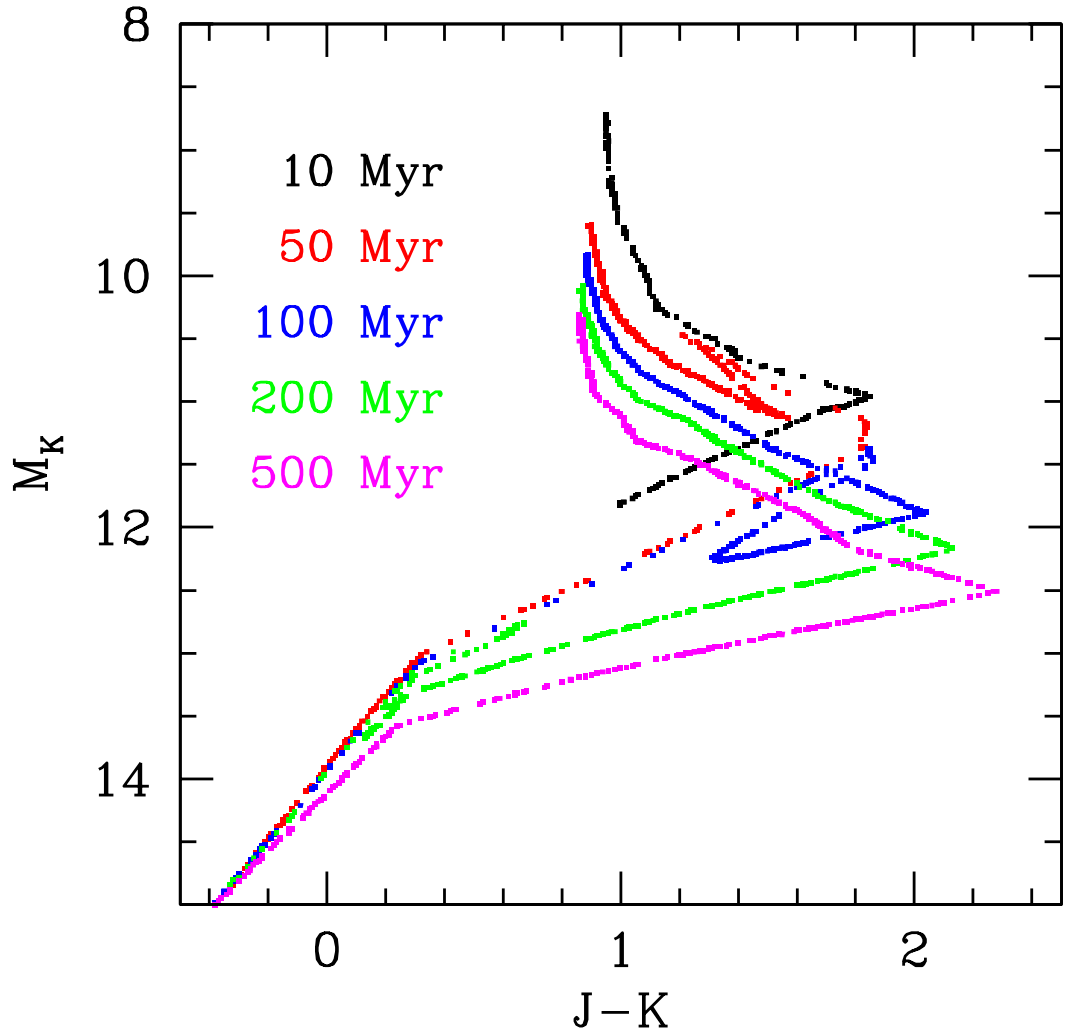}
   \caption{Synthetic color-magnitude diagram (MKO system) of cluster brown dwarf population shown in Fig. \ref{fig:cluster_hybrid}.
            Each sequence corresponds to a different age: 10$\,$Myr (black), 50$\,$Myr (red),
           100$\,$Myr (blue), 200$\,$Myr (green), and 500$\,$Myr (magenta), from top to bottom, respectively.  
           [{\it See the electronic edition of the Journal for a color version of this figure.}]}
    \label{fig:CMD_clusters}
\end{figure}
\clearpage

\begin{figure}
   \plotone{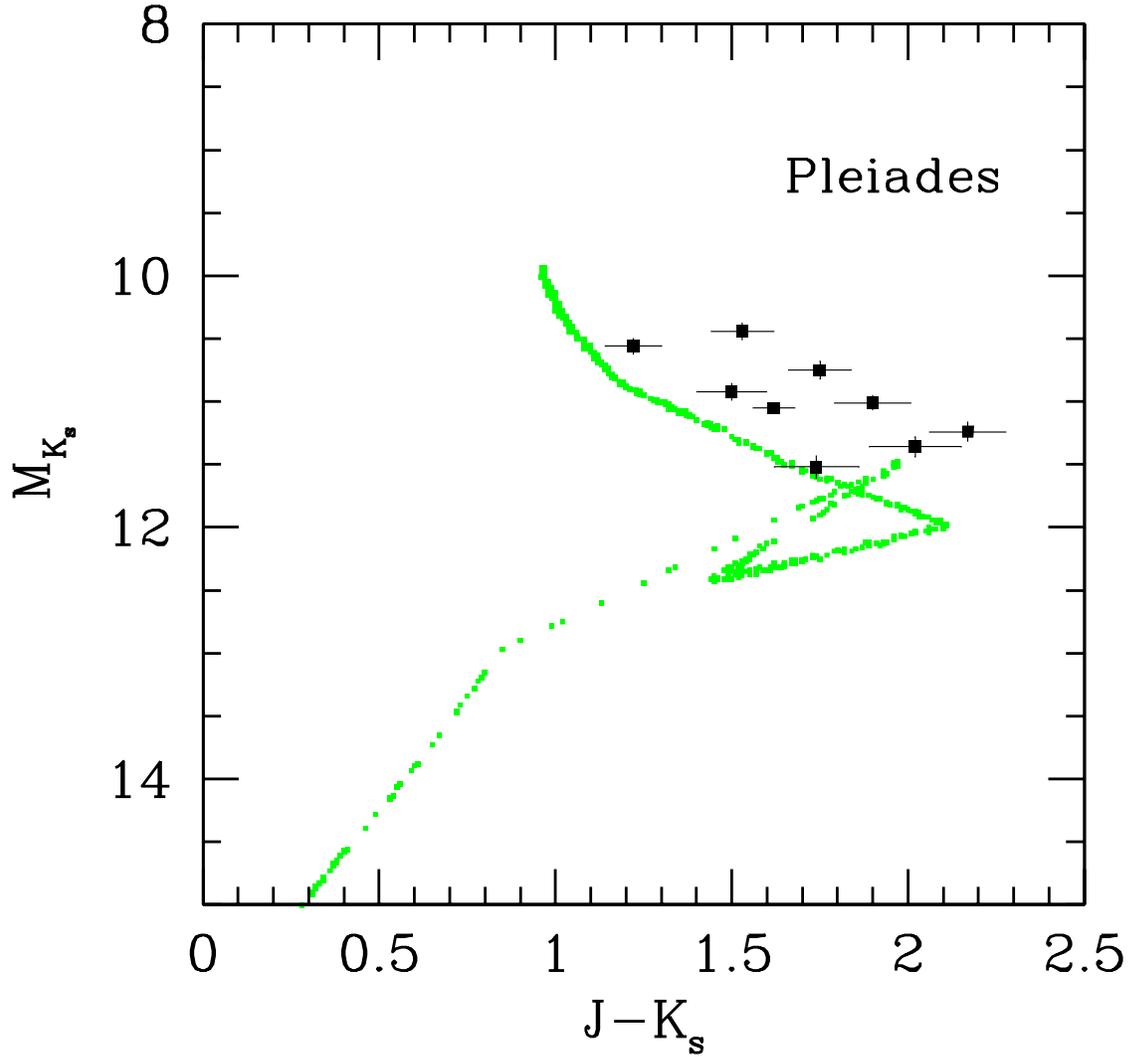}
   \caption{Color-magnitude diagram of the Pleiades.  A brown dwarf population based on the hybrid evolution sequence and colors (see text)
            with an IMF index of $\alpha=0.6$ and an age of 110$\,$Myr with a $\pm 5\,$Myr dispersion is shown in green.  The black
            squares show the 2MASS photometry reported in \citet{bihain06} for brown dwarf candidates with cluster
            membership confirmed by proper motion measurements.
           [{\it See the electronic edition of the Journal for a color version of this figure.}]}
    \label{fig:CMD_pleiades_bihain}
\end{figure}
\clearpage

\begin{figure}
   \plotone{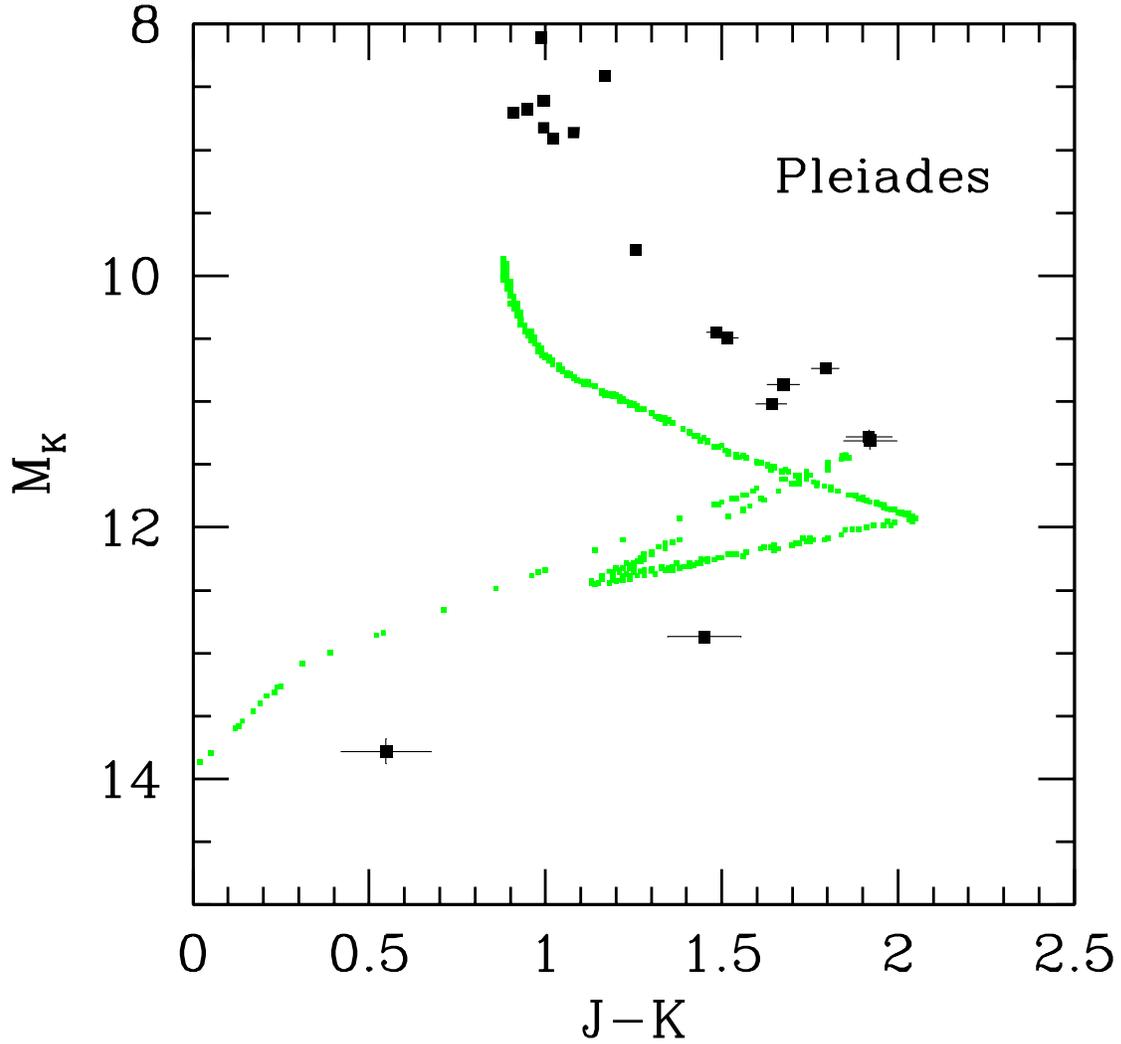}
   \caption{Color-magnitude diagram of the Pleiades.  A brown dwarf population based on the hybrid evolution sequence and colors (see text)
            with an IMF index of $\alpha=0.6$ and an age of 110$\pm 5\,$Myr is shown in green.  The black
            squares show the MKO photometry reported in \citet{casewell07}.
           [{\it See the electronic edition of the Journal for a color version of this figure.}]}
    \label{fig:CMD_pleiades_casewell}
\end{figure}
\clearpage

\begin{figure}
   \plotone{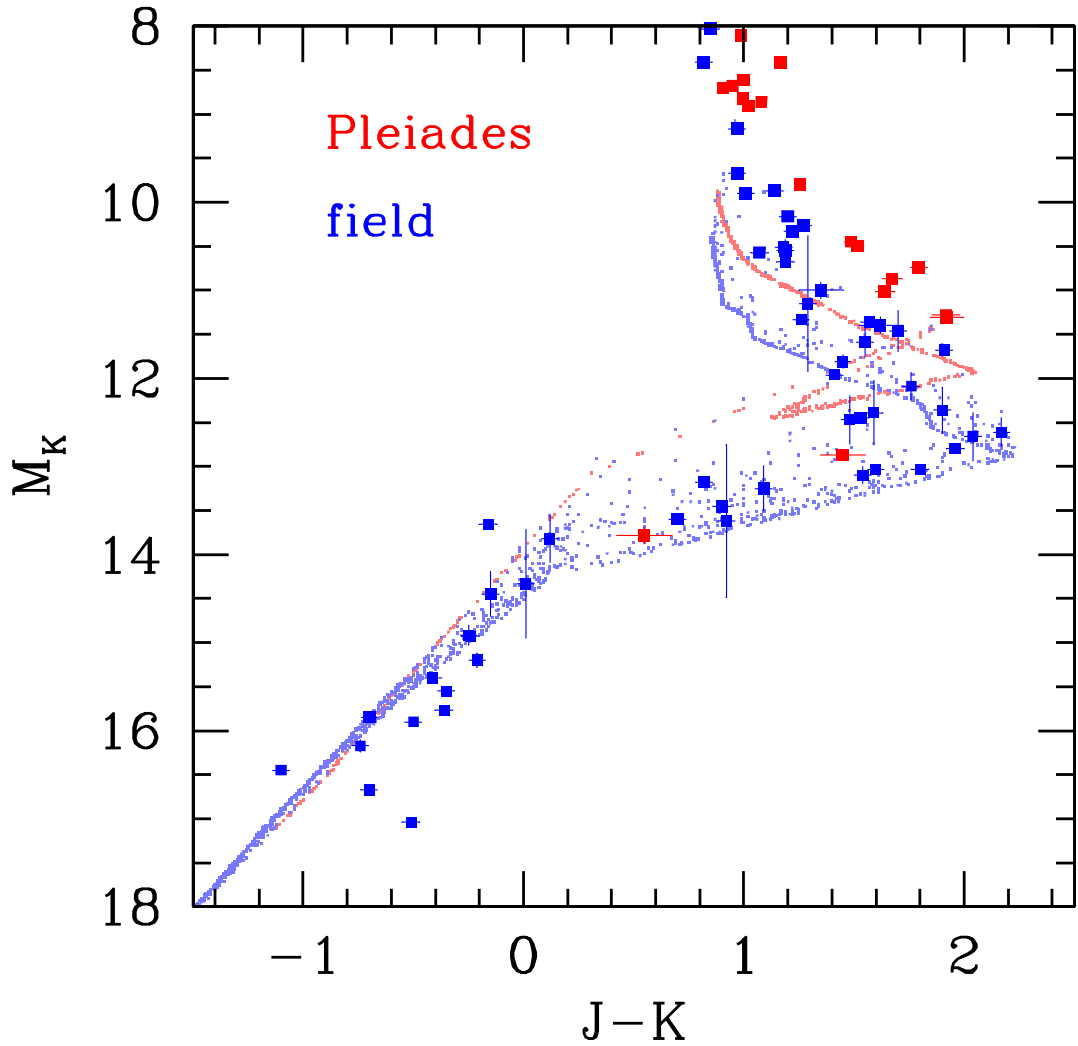}
   \caption{Color-magnitude diagram (MKO system) showing brown dwarfs in the Pleiades (red, \citet{casewell07}), and
            in the field (blue, see the caption of Fig. \ref{fig:CMD_data} for references).
            The corresponding hybrid model sequences are shown by small dots in lighter colors.
           [{\it See the electronic edition of the Journal for a color version of this figure.}]}
    \label{fig:all_data}
\end{figure}
\clearpage
\end{document}